\documentclass[12pt]{article}

\usepackage{amscd,amsmath,amssymb,amsfonts,xspace,mathrsfs}
\usepackage{graphicx}
\usepackage{fancyhdr}
\usepackage{rotating}
\usepackage{psfrag}
\usepackage{epsfig}
\usepackage{url}
\usepackage[unicode]{hyperref}
\usepackage{color}

\allowdisplaybreaks
\numberwithin{equation}{section}

\def\lfig#1#2#3#4#5{
\begin{figure}
\centerline{\hfill \includegraphics[height=#3]{#2}\hfill}
 \vspace{#5}
\caption{#1 \label{#4}}
\end{figure}}

\newcounter{tabl}
\newcommand{\be}{\begin{equation}}
\newcommand{\ee}{\end{equation}}
\def\ba{\begin{align}}
\def\ea{\end{align}}
\newcommand{\beq}{\begin{eqnarray}}
\newcommand{\eeq}{\end{eqnarray}}
\newcommand{\bea}[2]{\be\label{#2}\begin{array}{#1}}
\newcommand{\eea}{\end{array}\ee}

\newcommand{\bR}{\mathbb{R}}

\def\det{\,{\rm det}\, }
\def\diag{{\rm diag}}

\def\({\left(}
\def\){\right)}
\def\[{\left[}
\def\]{\right]}
\def\p{\partial}
\newcommand{\de}{\mathrm{d}}
\newcommand{\I}{\mathrm{i}}

\def\11{1\!\! 1}

\def\hf{\frac{1}{2}}

\def\d{\delta}

\def\eps{\varepsilon}

\def\l{\lambda}
\def\m{\mu}
\def\n{\nu}

\def\r{\rho}

\def\s{\sigma}

\def\om{\omega}

\def\L{\Lambda}

   \def\CC {{\cal C}}
   \def\CD {{\cal D}}

   \def\CG {{\cal G}}
   \def\CH {{\cal H}}

   \def\CL {{\cal L}}
   \def\CM {{\cal M}}
   \def\CN {{\cal N}}
   \def\CO {{\cal O}}

   \def\CS {{\cal S}}

   \def\CX {{\cal X}}

\newcommand{\hlam}{\hat\lambda}
\newcommand{\hmu}{\hat\mu}
\newcommand{\hnu}{\hat\nu}

\newcommand{\tE}{\lefteqn{\smash{\mathop{\vphantom{<}}\limits^{\;\sim}}}E}
\newcommand{\tP}{\lefteqn{\smash{\mathop{\vphantom{<}}\limits^{\;\sim}}}P}

\newcommand{\Et}{\lefteqn{\smash{\mathop{\vphantom{\Bigl(}}\limits_{\sim}
\atop \ }}E}

\newcommand{\tp}{\lefteqn{\smash{\mathop{\vphantom{\scriptstyle{c}}}\limits^{\sim}}}p}

\newcommand{\tN}{\lefteqn{\smash{\mathop{\vphantom{\Bigl(}}\limits_{\sim}
\atop \ }}N}
\newcommand{\tM}{\lefteqn{\smash{\mathop{\vphantom{\Bigl(}}\limits_{\,\sim}
\atop \ }}M}
\newcommand{\tNn}{\lefteqn{\smash{\mathop{\vphantom{\Bigl(}}\limits_{\,\sim}
\atop \ }}{\cal N}}

\newcommand{\ukap}{\lefteqn{\smash{\mathop{\vphantom{\ \atop \cdot}}\limits_{\!\sim}\atop\ }}\kappa}

\newcommand{\nd}{{\cal N}}

\newcommand{\Ref}[1]{(\ref{#1})}

\newcommand{\f}{\frac}

\def\xc{{\rm x}}
\def\yc{{\rm y}}

\def\xp{x_+}
\def\xm{x_-}
\def\xpm{x_\pm}

\def\xpd#1{x_{+,#1}}
\def\xmd#1{x_{-,#1}}

\def\xcp{\xc^+}
\def\xcm{\xc^-}
\def\xcpm{\xc^\pm}

\def\chif{\hat\chi}

\def\LMsim{\zeta}

\def\gind{q} 

\newcommand{\bC}{\mathbb{T}}
\newcommand{\bB}{\mathbb{B}}
\newcommand{\bG}{\mathbb{G}}
\newcommand{\eG}{G}
\def\Gn{R}
\def\Ln{L}
\def\Gg{\CC}
\def\Ggf{\Gg_\flat}

\def\lamgf{\lambda^\flat}

\def\Psigl{\Psi_0^{\rm gl}}

\def\zmcon{\Sigma}

\def\Ss{\CS}

\begin{document}

\title{\bf First order gravity on the light front}

\vspace{0.7cm}

\author{Sergei Alexandrov$^{a}$ and Simone Speziale$^b$ }

\date{}

\maketitle

\vspace{-1cm}
\begin{center}
$^a${\it Laboratoire Charles Coulomb, CNRS UMR 5221,
Universit\'e Montpellier 2, F-34095 Montpellier, France}\\
$^b${\it \small{Centre de Physique Th\'{e}orique, CNRS-UMR 7332, Luminy Case 907, 13288 Marseille, France}}
\end{center}
\vspace{0.1cm}
\begin{abstract}
We study the canonical structure of the real first order formulation of general relativity on a null foliation.
We use a tetrad decomposition which allows to elegantly encode the nature of the foliation
in the norm of a vector in the fibre bundle.
The resulting constraint structure shows some peculiarities.
In particular, the dynamical Einstein equations propagating the physical degrees of freedom appear
in this formalism as second class tertiary constraints, which puts them on the same footing as the Hamiltonian constraint
of the Ashtekar's connection formulation.
We also provide a framework to address the issue of zero modes in gravity, in particular, to study the non-perturbative
fate of the zero modes of the linearized theory.
Our results give a new angle on the dynamics of general relativity
and can be used to quantize null hypersurfaces in the formalism of loop quantum gravity or spin foams.
\end{abstract}

\newpage

\tableofcontents

\section{Introduction}
\label{sec-intro}

Null hypersurfaces play a pivotal role in the physical understanding of general relativity,
from the characterisation of gravitational radiation and exact solutions
\cite{Sachs:1962zzb,Bondi:1962px,Newman:1961qr,Scherk:1974zm,Frittelli:1995ed},
to the structure of isolated horizons and black holes \cite{Ashtekar:2004cn}.
Therefore, it is natural to ask whether one gets some interesting results if they are used
for the canonical formulation, namely, if one performs the 3+1 decomposition
and the canonical analysis with respect to a foliation of spacetime which
is not space-like, as usual, but light-like or null. This is the idea of the {\it light front} approach,
which has been put forward by Dirac \cite{Dirac:1949cp} and has been extensively developed in the context
of QCD and field theories in Minkowski spacetime leading to interesting results
in describing their quantum properties (see \cite{Burkardt:1995ct,Harindranath:1996hq,Grange:1998gy} for reviews).

In the context of gravity less has been done in this direction,  although already in his pioneering work \cite{Sachs:1962zzb}
Sachs showed that using a \emph{double-null} foliation the constraints imposing diffeomorphism invariance simplify,
and constraint-free data can be accessed as the conformal structure of the two-dimensional space-like metric embedded in the hypersurface.
This remarkable feature could in principle be used to reduce the canonical
dynamics to physical degrees of freedom only, which would obviously have
tremendous impact for both the classical theory and quantization attempts.

Partial success using a null foliation in general relativity is hindered by the more complicated
canonical structure caused by the fact that the induced metric on a null hypersurface
is degenerate. In particular, there is no natural projector nor induced affine connection.
One way to address this difficulty is to use the double-null foliation of Sachs,
which was promoted later into a $2+2$ formalism \cite{d'Inverno:1980zz}, where one picks
up two independent null directions and foliates spacetime by the two-dimensional space-like surfaces orthogonal
to both directions.
In this framework the Hamiltonian formulation in metric variables was carried out in \cite{Torre:1985rw}.
Its key feature is that the Hamiltonian constraint is second class, and does not generate gauge symmetries.
This can be intuitively understood because the condition that the hypersurface is null acts as a gauge-fixing condition,
and is consistent with the fact that there are no local infinitesimal deformations
mapping a null hypersurface into a neighbouring null hypersurface.
The presence of second class constraints makes the canonical formulation quite complicated,
and neither the reduced phase space has been constructed, nor the Dirac brackets explicitly
evaluated, revealing the symplectic structure to be quantized.

The canonical analysis of general relativity is simplified using Ashtekar variables \cite{Ashtekar:1986yd,Ashtekar:1987gu},
that is a densitized co-triad and a self-dual Lorentz connection.
The light front formulation in Ashtekar variables was constructed in \cite{Goldberg:1992st,Goldberg:1995gb}
and further investigated using the 2+2 formalism in \cite{d'Inverno:1995,d'Inverno:2006hr,d'Inverno:2006dy}.
These formulations expose additional features of the light-front theory, including the nice property
that the first class part of the constraint algebra forms
a Lie algebra, with proper structure constants, given by the semi-direct product
of the hypersurface diffeomorphisms and the internal symmetry group.
However, a difficulty with this approach is posed by the reality conditions
needed for Lorentzian signature \cite{Bengtsson:1989pe,Immirzi:1992ar}.
These conditions become especially problematic at the quantum level where no consistent way of implementing them has been found
so far.\footnote{See \cite{Ashtekar:1995qw,Alexandrov:2005ng,Wieland:2010ec} for some attempts in this direction.}
A way to avoid this complication is to work with real connection variables, as it is
done in the modern approaches to quantum gravity via loop and spin foam techniques
\cite{Rovelli:2004tv,Alexandrov:2010un,Perez:2012wv}.
Therefore, it would be desirable to extend the previous Hamiltonian analysis of general relativity
on a null foliation to such real formulation. This is precisely the goal of this paper.

Such extension is useful for several reasons. Firstly, to contribute a new attack line
to the problem of finding a reduced canonical formalism in terms of physical
degrees of freedom only. Secondly, to analyse the initial value problem and identifying
the constraint-free data in terms of a real Lorentz connection.
Finally, framing the theory in these variables would make it possible to try a light front
quantization of gravity using the techniques of loop quantum gravity or spin foams,
in particular, defining a dynamics for the null twisted geometries introduced in \cite{Speziale:2013ifa}.

With these motivations in mind, in this paper we study the light front formulation
of general relativity in the first order tetrad formalism, where the Einstein action takes the following form\footnote{Notice
that we chose units $8\pi G=1$, instead of the more common choice of normalising $16\pi G$.
This is in order to avoid a number of factors of 2 in the canonical analysis.}
\be
S[e,\om]=\f{1}{4}\int_\mathcal{M}\eps_{IJKL} e^I\wedge e^J \wedge \(F^{KL}(\omega)-\frac{\Lambda}{6}\, e^K\wedge e^L\).
\label{HPaction}
\ee
Here $e^I$ is the tetrad 1-form and
$F^{IJ}(\omega)=\de\omega^{IJ}+{\omega^{I}}_K\wedge \omega^{KJ}$ is the curvature
of the spin connection $\om^{IJ}$.
The canonical analysis of this action using a space-like foliation can be found, for instance, in \cite{Peldan:1993hi}
(for the analysis of tetrad gravity in the second order formalism, see \cite{Deser:1976ay}).
On the other hand, the canonical analysis on a null foliation has not been studied before
and we fill this gap here.\footnote{In \cite{Aragone:1989py} the authors do study
the action \eqref{HPaction} on the light front, but they do not perform its canonical analysis.}
The immediate advantage of working with tetrads is that one can use the standard $3+1$ splitting
and reproduce all features of the $2+2$ formalism used in the literature from the natural
double-null foliation of the Minkowski metric in the fibre space.
In particular, the nature of the foliation can be controlled by the norm of an internal space vector,
and in the null case one can describe the degenerate induced metric on the hypersurface
while keeping the triad invertible, a property which makes the canonical
variables and calculations more transparent.

Our first result is to fully characterize the system of constraints, and to show that
the reduced phase space has two dimensions per point of the null hypersurface, consistently with
two local degrees of freedom of gravity.\footnote{This counting on the light front
may be unfamiliar to some readers, and it is explained in section \ref{sec-LFfeatures}.}
As in other light front formulations, the Hamiltonian constraint is second class, whereas first class constraints
generate a genuine Lie algebra given by the semi-direct product of the spatial diffeomorphisms
and the internal gauge group associated with the isometries of a null hyperplane.
The system possesses secondary constraints, familiar to people working with Plebanski formulation of general relativity
and capturing a part of the torsionless condition in the canonical framework, but also two tertiary constraints.
These are shown to be precisely the dynamical equations propagating the two physical degrees of freedom.
The fact that dynamical equations are turned into constraints is a unique feature of
combining a first order formalism with a null foliation.

Our analysis also sheds light on a few other issues.
In particular, the gauge fixing used to write the action on the light front
leads  to the apparent loss of one field equation.
This issue has been dealt with
by either adding the missing equation by hand \cite{Torre:1985rw},
or by extending the phase space and slightly modifying the action \cite{Goldberg:1992st}.
We demonstrate that in the first order formalism the apparently missing equation
is automatically obtained via the stabilisation procedure.
Thus, the original action contains all of Einstein's equations, and no modification like those proposed
in the literature is needed.

Another important issue which we discuss concerns zero modes.
As is well known from the light-front analysis of field theories in Minkowski spacetime,
specifying a unique solution in the light front formalism may also require, on top of initial conditions
of the physical fields, some additional data in the form of their zero modes.
This issue becomes especially pressing at quantum level, where the zero modes are expected
to carry non-trivial properties of the vacuum. To the best of our knowledge,
it has not been tackled before in the literature on general relativity, and we address it here for the first time.

Except for the analysis of zero modes, we restrict our attention to local considerations. In particular, key dynamical questions
such as the actual extension of the null sheet before caustics form, the analysis of boundary and asymptotic conditions,
or the inclusion of matter will be discussed in future work. In practice, this means that we allow ourselves
to perform integration by parts, and neglect boundary terms.

The paper is organized as follows.
First, we recall some features of the light front field theories which might be unfamiliar to
some of the researchers working in general relativity and quantum gravity.
Then in section \ref{sec-decomp} we introduce the 3+1 decomposition, formulate the condition ensuring that
the foliation is light-like, and analyze how this affects the nature of the Lagrangian equations of motion.
The canonical analysis is presented in section \ref{sec-canonical} where we find and classify all constraints
of the theory. Next, in section \ref{sec-pecular} we discuss various peculiarities of the resulting formulation.
Finally, section \ref{sec-concl} is devoted to conclusions. Few appendices contain additional helpful information.
Thus, in appendix \ref{ap-scalar} we review the light front formulation of a scalar field theory. In appendix \ref{ap-metric}
we provide explicit expressions for the inverse tetrad and the metric induced by our 3+1 decomposition. Appendix \ref{ap-algebra}
presents explicit results for the constraint algebra used in the course of our canonical analysis. And finally in appendix
\ref{ap-lost} we analyze the effect of a gauge fixing of constraint systems on their Lagrangian equations of motion.

According to our conventions, the metric has mostly plus signature. In particular, the flat Minkowski metric is
$\eta_{IJ}=\diag(-1,1,1,1)$. Components of spacetime tensors are labeled by greek indices $\mu,\nu,\dots=0,1,2,3$.
Their spatial components are labeled by latin indices from the beginning of the alphabet $a,b,\dots =1,2,3$.
Components of the tangent space tensors are labeled by capital latin indices $I,J,\dots =0,1,2$,
whereas their spatial parts are labeled by latin indices from the middle of the alphabet $i,j,\dots =1,2,3$.
The Levi-Civita symbol with flat indices is normalized as $\eps_{0123}=1$.
On the other hand, for the antisymmetric tensor density with spacetime indices
we set $\eps^{0abc}=\eps^{abc}$. This opposite sign convention avoids a cumbersome minus sign in the definition
of the determinant of the tetrad. Finally, the symmetrisation and anti-symmetrisation
of indices is denoted respectively by {\footnotesize{$(\cdot\, \cdot )$}} and {\footnotesize{$[\cdot\,\cdot]$}},
and includes the normalisation weight $1/2$.

\section{Generic features of the light front formalism}
\label{sec-LFfeatures}

Before considering the gravitational case, where null hypersurfaces are dynamical,
we would like to recall some generic features of the light front formalism in Minkowski spacetime,
which will be of help in understanding the gravitational case.
In particular, it will allow us to highlight the presence of zero modes and the role they play,
and the way degrees of freedom are counted in the canonical framework.

The idea of using a null foliation for the canonical analysis dates back to Dirac \cite{Dirac:1949cp}.
He suggested to introduce light-front coordinates in one of the Lorentzian planes, for example
\be
\xcpm=\frac{1}{\sqrt{2}}\(\xc^0 \pm \xc^3\),
\label{lfcoor}
\ee
and to consider one of them, say $\xcp$, as the time coordinate for the canonical analysis.
A distinguishing property of such choice is that the hypersurfaces $\xcp=const$ have maximal number of isometries:
because the induced metric has one degenerate direction, the isometry group has seven generators,
as opposed to the six generators for a space-like hypersurface $\xc^0=const$.

This fact makes field theories on the light front very specific.
Indeed, some peculiarities can be noticed already from the mass shell condition.
Taking as an example the case of a scalar field theory, one finds that in coordinates \eqref{lfcoor}
it becomes a {\it linear} equation for the momentum variable $p^-$ playing the role of the energy in the light front frame
\be
p^- = \frac{(p^\bot)^2+m^2}{2p^+}.
\label{masshell}
\ee
As a consequence, the physical vacuum is always {\it trivial} and coincides with the state with vanishing energy-momentum $P^\mu=0$.
Indeed, whereas in the conventional approach the vacuum is modified by interactions
and the true vacuum can be a state with non-vanishing energy, on the light front
the relation \eqref{masshell} implies that any physical state of a (massive) particle
must have positive longitudinal momentum $p^+>0$. Hence, a physical vacuum with a non-zero energy $P^-$
and vanishing momentum cannot exist.

The triviality of the vacuum is a tremendous technical advantage, and many of the successes of light front quantization derive from it.
However, it raises the question of how non-perturbative effects such as spontaneous symmetry breaking can be incorporated.
It turns out that such non-trivial effects are hidden in the {\it zero mode} sector of the theory, describing the modes $\phi_0$
with vanishing longitudinal momentum $p^+$, or in other words, satisfying $\p_-\phi_0=0$ \cite{Yamawaki:1998cy}.
The special role of these modes is clearly seen already in the mass shell condition \eqref{masshell},
which is ill-defined at $p^+=0$. Regularising this divergence requires in turn
a careful choice of boundary conditions at $\xcm\to \pm \infty$, see e.g. \cite{Heinzl:1993px}.
Boundary conditions effectively play a subtle role in the light front formalism,
as different choices may lead to different physical results via the change in the dynamics of the zero mode sector.

Another generic feature of the light front formalism, which is more directly relevant for the present paper,
is the appearance of constraints in the canonical analysis.
To see how they arise, it is sufficient to consider the standard kinetic term for the scalar field $\phi$.
Picking $\xc^+$ as a time variable, it becomes {\it linear} in the ``velocities" $\p_+\phi$, since
\be
\hf\((\p_0\phi)^2-(\p_3\phi)^2\)=\p_+\phi\p_-\phi.
\ee
Hence, if the interaction does not depend on derivatives of the field, the conjugate momentum
$\pi:=\d {\cal L}/\d \p_+\phi$ is independent of velocities, and one gets the constraint
\be
\Psi:= \pi-\p_-\phi=0.
\label{lfcon}
\ee
Furthermore, it is easy to check that this constraint is second class since it does not commute with itself,
\be
\{\Psi(\xc),\Psi(\yc)\}=-2\p_{\xc^-}\delta^{(3)}(\xc-\yc).
\label{lfcom}
\ee
This in turn implies that the field itself is non-commutative and the correct symplectic structure is given by a Dirac bracket,
with typical form
\be
\{\phi(\xc),\phi(\yc)\}_D=\Delta^{-1}(\xc,\yc),
\label{lfcomphi}
\ee
where $\Delta$ is the operator on the r.h.s. of \eqref{lfcom}.
In the momentum representation
the commutator is proportional to $1/p^+$ which gives rise to the same divergence as the one appearing
in the mass shell condition \eqref{masshell} showing again the special role of the zero modes.
In some theories the concrete form of the above constraints and commutators can be a bit different,
but the general mechanism remains essentially the same.
In particular, in the linearized approximation the physical modes of both gauge theories and gravity
satisfy exactly the same relation \eqref{lfcon}.

Notice that in the above example the momentum $\pi$ can be excluded by means of the light front constraint \eqref{lfcon}.
As a result, one gets a one-dimensional phase space described by the field $\phi$ only,
with the non-trivial symplectic structure given by the Dirac bracket \eqref{lfcomphi}.
Thus, the (infinite) dimension of the phase space matches the number of degrees of freedom,
without the usual factor 2 of the equal-time approach, and this conclusion
turns out to be valid for {\it any} theory on the light front.

Since we are talking about infinite-dimensional spaces,
there is actually nothing surprising that the $2n$-dimensional phase space of
one formulation can be packed into $n$-dimensions in the other.
It is nonetheless instructive to see explicitly how the mapping goes.
To that end, let us consider the decomposition of the field $\phi(\xc)$ into Fourier modes.
The standard decomposition reads
\be
\phi(\xc)=\int \de^3 \vec p \[ a(p) e^{\I p_i \xc^i-\I \omega \xc^0}+a^*(p) e^{- \I p_i \xc^i+\I \omega \xc^0}\],
\ee
where $\omega=\sqrt{\vec p^2+m^2}$. The presence of the two terms, or equivalently the complexity of $a(p)$,
explains the bi-dimensionality of the usual phase space.
On the other hand, in the light cone coordinates the decomposition is given by
\be
\phi(\xc)=\int \de p^+\de^2 p^\bot \[ b(p^+,p^\bot) e^{\I (p\cdot \xc)^\bot-\I p^+ \xcm -\I \, \frac{(p^\bot)^2+m^2}{2p^+}\, \xcp}\]
\ee
with $b(-p)=b^*(p)$. The presence of only one term, which corresponds to the one-dimensionality of the light front phase space,
can be traced back to the fact that, in contrast to the usual case, the linear mass shell condition \eqref{masshell}
restricts the spectrum of the light cone momentum to the half positive light cone.
The map relating the two decompositions is given by
\be
b(p^+,p^\bot)=a\(\frac{1}{\sqrt{2}}\(p^+ -\frac{(p^\bot)^2+m^2}{2p^+}\),p^\bot\),
\qquad
p^+ >0.
\ee
and maps the positive half-axis of $p^+$ into the whole real line of $p^3$.

Finally, let us go back to the issue of zero modes.
As we discuss in Appendix A, generically at classical level the zero modes turn out to be determined by the field equations appearing
as additional second class constraints.
However, a special situation arises for massless theories, and this can be understood easily on physical grounds.
The particularity of this case can be seen from the fact that at each point in spacetime there is a particle worldline which
is parallel to the light front hypersurface (see Fig. \ref{fig-cone}). Thus, it never intersects the initial value
surface of the light front formulation and therefore is not determined by the initial data.
We call the corresponding modes of the fields by {\it global zero modes}
as they have vanishing momenta $p^+=p^\bot=0$ or, equivalently, are independent of all hypersurface coordinates.

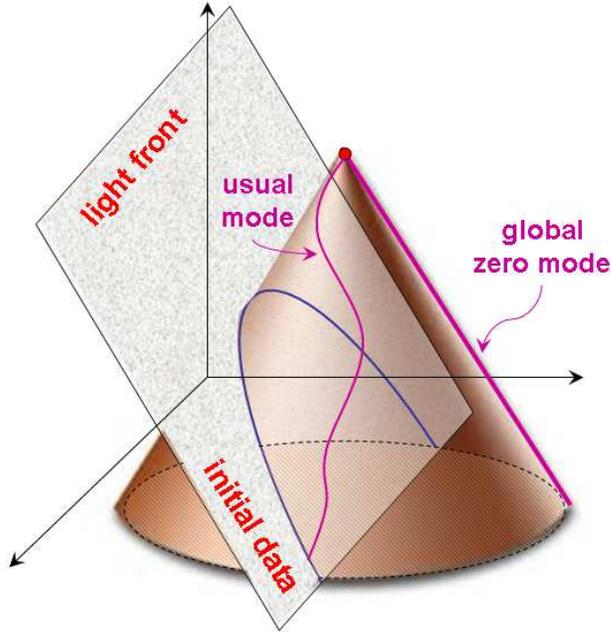
\lfig{\small{\emph{The past light cone of an event in spacetime. All world-lines intersect
the light front hypersurface in a finite time, except the one parallel to it.
The latter corresponds to the global zero mode.}}}{cone}{10cm}{fig-cone}{-1.3cm}

It is clear that to uniquely determine the evolution, the global zero modes should be supplemented to the initial data.
It turns out that the canonical formulation ensures this in an interesting way:
in the massless theory, the global zero mode of the light front constraint \eqref{lfcon} is first class, and the corresponding
undetermined Lagrange multiplier provides the additional missing data.
In the massive case, the zero mode is instead converted into second class by the presence of another constraint,
which is the one imposing an equation on the zero mode of the field itself and making
the initial value problem on the light front well defined. For the interested reader,
details of this constraint analysis are reported in Appendix \ref{ap-scalar}, including the special two-dimensional case
where the two notions of zero modes obviously coincide.

\section{Null foliation via tetrads}
\label{sec-decomp}

We set up the canonical analysis following the standard ADM formalism for a space-like 3+1 splitting.
We assume the spacetime manifold to be of the form $\CM=\bR\times\Sigma$
where $\Sigma$ is non-compact\footnote{Boundary conditions do play a non-trivial role on the light front. We will comment on this below.},
and take adapted coordinates $\xc^\m = (t, \xc^a)$.
However, in contrast to the ADM formulation, we choose the level sets of the time parameter to be null hypersurfaces.

This choice has an immediate consequence on the theory.
The crucial difference between general relativity and field theories is
that the metric and thus the causal structure of spacetime are dynamical.
Therefore, whereas in field theories the choice of a null foliation is merely a choice of coordinates,
in general relativity this is a (partial) gauge fixing condition:
requiring that the level sets of a coordinate $t$ are null fixes one of the metric components,
\be
g^{00}=g^{-1}(\de t,\de t)=0.
\label{lf-gauge}
\ee
Thus, gravity on the light front is a partially gauge-fixed theory.\footnote{Implementing the gauge fixing
in the action leads to the apparent loss of the Einstein equation corresponding to the variation
of the action with respect to $g^{00}$. This was noticed in \cite{Torre:1985rw,Goldberg:1992st}
and it was suggested to modify the gauge-fixed action as
to restore the `lost' equation. However, as we will show in section \ref{subsec-lost},
no modification is necessary in the first order formalism, as the desired equation is obtained by means of the stability conditions.}

This gauge implies that the leaves $\Sigma$ are null, which means that their induced metric is degenerate
and there is no natural affine structure.
It turns out that these technical difficulties can be elegantly dealt with using tetrads.
Below in this section we show how the gauge \eqref{lf-gauge} can be nicely implemented in the tetrad formalism
and which consequences it implies on the Lagrangian equations of motion derived from the action \eqref{HPaction}.

\subsection{Decomposition of the tetrad}
\label{subsec-dectetrad}

Our starting point is the general tetrad decomposition introduced in \cite{Alexandrov:1998cu},
and used in the Lorentz covariant approach to loop quantum gravity \cite{Alexandrov:2000jw,Alexandrov:2002br},
\be
e^0=N \de t+\chi_i E_a^i\de x^a,
\qquad
e^i=N^a E_a^i\de t +E_a^i\de x^a.
\label{docomptet}
\ee
Here $E_a^i$ is the triad, $N$ and $N^a$ are related to the lapse and shift functions,
and $\chi^i$ describes the remaining 3 components of the tetrad.
This decomposition generalises the one commonly used in the canonical analysis in tetrad variables,
which is adapted to the ADM variables by aligning $e^0$ with $\de t$ and thereby setting $\chi^i=0$, a choice referred to as ``time gauge"
in the literature.
A drawback of this generalization is that, not being adapted to the coordinates,
$N$ and $N^a$ do not coincide with the lapse and shift Lagrange multipliers. Instead, as we explain below,
they are related to them by a linear transformation.
On the other hand, it is the introduction of the additional variables $\chi^i$
that allows us to put the theory on the light front in an elegant way.
The reason is that $\chi^i$ controls the normal to the hypersurface $t=const$ and thus the foliation.
Equivalently, the hypersurface normal can be encoded in the following vector in the internal space with the flat Minkowski metric
\be
x^I=\(1, \chi^i\).
\label{unitvectors}
\ee
In particular, the norm of this vector controls the nature of the foliation: it is space-like, light-like, or time-like
if $x^2$ is less, equal, or larger than 0. To see this, it is sufficient to look at the induced metric on $\Sigma$,
which is found to be
\be
\gind_{ab} := \CX_{ij} E_a^iE_b^j, \qquad  \CX_{ij} := (\delta_{ij}-\chi_i\chi_j).
\label{projchi}
\ee
It has the signature $(+++)$, $(0++)$ or $(-++)$ in the above three cases, respectively.
Alternatively, one can compute the inverse metric obtained from \eqref{docomptet},
which gives $g^{00} \propto \chi^2-1$.
The fact that changing $\chi^2$ we can change the type of the foliation allows to describe all of them in a uniform way.
For instance, in \cite{Alexandrov:2005ar} the decomposition \eqref{docomptet} was used to get the spectrum
of the area operator in loop quantum gravity for time-like surfaces. Here we are rather interested in the light-like case,
which in terms of the variables introduced by this decomposition reads
\be
\chi^2=1
\qquad \mbox{or}\qquad x^2=0.
\label{lfcond}
\ee
Thus, the light front condition \eqref{lf-gauge} becomes a condition on the norm of the internal space vector.
When it holds, the matrix $\CX_{ij}$
becomes a projector, and so does $\gind_{ab}$, with the null eigenvector given by
\be\label{nullvec}
\p_- = E^a_i\chi^i\p_a.
\ee

Note that, despite the degeneracy of the induced metric \eqref{projchi} for $\chi^2=1$, the triad
$E^i_a$ can always be assumed to be invertible, and used to map hypersurface indices to internal indices.
We see this as another advantage of our formalism.
The inverse triad will be denoted as usual by $E^a_i$,
and allows us to define the induced metric with mixed and contravariant indices,
\be
\gind^{a}_{b}=\CX^i_jE^a_iE_b^j,
\qquad
\gind^{ab}=\CX^{ij}E^a_iE^b_j.
\ee
The latter should not be mistaken with a sub-matrix of $g^{\m\n}$, whose expression is reported explicitly in \Ref{invmetric}.
Furthermore, we use the triad determinant $\sqrt h := \det E^i_a \neq 0$ to define tensor densities,
and, as customary in the literature, we keep track of the density weights using tildes, e.g.
\be
\tE^a_i = \sqrt h E^a_i, \qquad \tN =  \f1{\sqrt h}\, N.
\ee

A pay-off for the universality of \Ref{docomptet} is that it is not adapted to the choice
of coordinates.
Due to this, the functions $N$ and $N^a$ there appearing are not immediately the lapse and
shift functions which solder the $3+1$ decomposition and appear as Lagrange multipliers
for the Hamiltonian and the vector constraints in the decomposition of the action.
One way of establishing the relation between them is to compute the metric associated with \eqref{docomptet},
\be\label{metric1}
g_{\mu\nu} = \(\begin{array}{cc}
-N^2 + E^i_a E^i_b N^a N^b  & E^i_b E^i_c N^c - N E_b^i\chi_i
\\
E^i_a E^i_c N^c - N E_a^i\chi_i & \gind_{ab}
\end{array}\),
\ee
and to find the linear change of variables that puts it in the ADM form.
For generic $\chi^i$, this is achieved via \cite{Alexandrov:1998cu}
\be
N=\CN+E_a^i\chi_i\nd^a, \qquad N^a=\nd^a+E^a_i\chi^i\CN,
\label{redNN}
\ee
where $\CN$ and $\CN^a$ are the proper lapse and shift functions. This is confirmed also by the canonical analysis,
which identifies them as the Lagrange multipliers of the diffeomorphism constraints.
However, the redefinition \Ref{redNN} is singular for $\chi^2=1$, which means that in the light-like case
the metric cannot be put in the ADM form.
This is again a consequence of the lack of a natural projector on a null hypersurface.
Nonetheless, it is possible to identify the canonical lapse function $\CN$ by computing the determinant of the tetrad, which gives
\be
e = \(N-E_a^i\chi_i N^a\)\sqrt{h}.
\ee
This suggests to define the lapse via the same transformation \eqref{redNN} as for generic $\chi^i$,
\be
N=\CN+E_a^i\chi_i N^a,
\label{deflapse}
\ee
As we show below, this definition matches the identification of the lapse as a Lagrange multiplier.
On the other hand, there is no canonical definition for the shift vector.\footnote{As it will be
clear below when we present the Hamiltonian form of the action, the existence of a canonical choice for lapse,
and the arbitrariness of shift vector, is related to the fact that there is a canonical expression for the
constraint generating spatial diffeomorphisms, whereas we lack such an expression for the Hamiltonian constraint.
The reason is that the latter includes a projection of the curvature on the hypersurface $\Sigma$, but
such a projector cannot be defined in a unique way.}
We choose it to be simply $N^a$ as in the original decomposition \Ref{docomptet}.
Further details on the $3+1$ decomposition, such as expressions for the inverse tetrad,
the metric and its inverse can be found in appendix \ref{ap-metric}.

The tetrad formalism allows us to elegantly recover the $2+2$ formalism of \cite{d'Inverno:1980zz}.
To that end, observe that by taking the parity or time-reversed transform of $x^I$ we obtain
a pair of null vectors that foliate Minkowski spacetime via two-dimensional space-like planes. Denote this pair
\be
\xpm^I=(\pm 1, \chi^i).
\label{nullvecx}
\ee
Then one can easily write projectors on the double-null Minkowski foliation, and map them
to the tangent space via the tetrad. This operation provides us with a projector
${\perp^\m}_\n$ on the 2-dimensional space-like surface $\Ss$ contained in $\Sigma$
and its complement $\d^\m_\n - {\perp^\m}_\n$ projecting on the time-like surface spanned by the image of \Ref{nullvecx}.
Since $\Sigma$ is defined by the level sets of $t$, we have in particular that ${\perp^a}_b= q^a_b$.

\subsection{Field equations}
\label{subsec-Einstein}

Before performing the canonical analysis, it is useful to look at the effects of the light front condition
from the perspective of the covariant field equations. This will allow us to identify the splitting into
constraints and dynamical equations, in particular exposing the fact that with the gauge fixing \Ref{lf-gauge}
the lapse function is determined in terms of other fields.
The field equations obtained from \Ref{HPaction} read
\begin{align}
\bC_{\mu\nu}^I :=&\, D_{[\mu}e_{\nu]}^I=  0,
\label{Cartan}
\\
\bG^\m_I:=&\, \eG^\mu_I + \L e^\m_I = 0,
\label{Einst}
\end{align}
where $D=d+\om$ is the covariant derivative and $\eG^\m_I = (e^\m_K e^\r_I -\hf\, e^\m_I e^\r_K )  e^\s_L  F^{KL}_{\r\s}$
is the Einstein tensor.
The first set of equations is the torsion-free condition or Cartan (second structure) equation and, provided the tetrad is invertible,
it is uniquely solved by the Levi-Civita connection $\om^{IJ}(e)$. The 16 tetrad equations can be split into 10 equations
for the symmetric Einstein tensor, and 6 equations imposing the vanishing of the antisymmetric part of the Ricci tensor.
The latter vanishes automatically in the absence of torsion, thus reducing the field equations to Einstein's equations.

\subsubsection{Cartan equations}
\label{subsec-Cartan}

Let us look first at the Cartan equations imposing the vanishing of the torsion.
Using the 3+1 decomposition \eqref{docomptet}, the 24 equations \Ref{Cartan} split as follows,
\begin{subequations}
\beq
\bC_{ab}^i&=& \p_{[a}E_{b]}^i+\omega^{ij}_{[a}E_{b],j}+\omega^{0i}_{[a}E_{b]}^j\chi_j ,
\label{Ceq1}
\\
\bC_{ab}^0&=&\p_{[a}\(E_{b]}^i\chi_i\)+\omega^{0i}_{[a}E_{b],i},
\label{Ceq2}
\\
\bC_{0a}^i&=& \p_t E^i_a-\p_a\(N^b E_b^i\)+\omega_0^{ij} E_{a,j}
+\omega_0^{0i} E_a^j\chi_j-\omega_a^{ij}N^b E_{b,j}-\omega_a^{0i}\(\CN+N^b E_b^j\chi_j\),
\label{Ceq3}
\\
\bC_{0a}^0&=& \p_t\( E_a^i\chi_i\)-\p_a\(\CN+N^b E_b^i\chi_i\)+\omega_0^{0i}E_{a,i}-\omega_a^{0i} N^b E_{b,i}.
\label{Ceq4}
\eeq
\end{subequations}
The first two sets of equations, \eqref{Ceq1} and \eqref{Ceq2}, do not depend on time derivatives nor on $\omega_0^{IJ}$
and therefore they will be identified as $9+3=12$ constraints in the canonical theory. As we will see, they correspond
to 6 constraints related to the gauge transformations in the tangent space and to 6 secondary second class constraints.
The remaining equations \eqref{Ceq3} and \eqref{Ceq4} contain time derivatives,
and canonically are expected to correspond to Hamiltonian equations of motion.
However, combining \eqref{Ceq4} in the appropriate way with other equations, it is possible to obtain the following result
\be
E_a^i\Bigl[\p_t\chi_i+\omega^{ij}_{0}\chi_j+\CX_{ij} \omega_0^{0j}
-N^b\(\p_b\chi_i+\omega^{ij}_{b}\chi_j+\CX_{ij} \omega_b^{0j}\)\Bigr] -\p_a\CN+ \CN\omega_a^{0i}\chi_i=0.
\label{rewCeq4}
\ee
Contracting \eqref{rewCeq4} with $E^a_i\chi^i$ and imposing the light front condition \eqref{lfcond}, one finds
\be
E^a_i\chi^i\(\p_a\log\CN-\omega_a^{0j}\chi_j\)=0.
\label{constrN}
\ee
This result shows that, whereas generically all 3 equations \eqref{rewCeq4} are Hamiltonian equations of motion,
precisely on the light front one of them becomes independent of time derivatives and
should rather be interpreted as an equation for the lapse function. Since the latter is the Lagrange multiplier
of the Hamiltonian constraint, this means that the constraint will be second class.
This is a well-known conclusion (see \cite{Torre:1985rw,Goldberg:1992st,d'Inverno:1995}), which
is consistent with the fact that in gravity the light front condition appears as a partial gauge fixing.
The above analysis shows that the symmetry which is gauge fixed corresponds to
the time diffeomorphisms generated by the Hamiltonian constraint.

Note that the equation \eqref{constrN} fixing the lapse is differential so that it does not fix $\CN$ uniquely.
The differential operator acting on the lapse is nothing else but $\p_-$.
Thus, the undetermined part of $\CN$ is the typical zero mode on the light front.
We will explain its appearance in section \ref{subsec-diff}.

Finally, it is useful to discuss what happens if one considers the theory where all components of the vector $\chi^i$
are taken to be not dynamical variables, but fixed functions. This means simply that one fixed the boost gauge freedom in the tangent space.
In this case, from the canonical point of view, the three equations \eqref{rewCeq4} are generically interpreted not as equations of motion,
but as equations on $\omega_0^{IJ}$ which play the role of the Lagrange multipliers.
This is in agreement with the gauge fixing of the three boosts which converts three first class constraints
into second class. However, as above, on the light front the interpretation changes.
The equation \eqref{constrN} does not depend not only on time derivatives, but also on $\omega_0^{IJ}$.
Thus, instead of three equations on $\omega_0^{IJ}$, one has two on $\omega_0^{IJ}$ and one on $\CN$.
As a result, we expect that on the light front only two constraints generating local Lorentz transformations in the tangent space
are converted into second class. This can be traced back nicely to
the fact that the stability group of a null surface has one more generator comparing to the space-like case.

\subsubsection{Tetrad equations}

Next, we turn to the tetrad equations \eqref{Einst}. They can be decomposed as
\begin{subequations}
\beq
e \bG^0_0 &=& -\frac14\,\eps_{ijk} \eps^{abc} E^i_a F^{jk}_{bc} +\Lambda \sqrt{h} ,
\label{Geq1}\\
e \bG^0_i &=& \frac14\,\eps_{ijk} \eps^{abc} \( E_a^l \chi_l F^{jk}_{bc} - 2 E^j_a F^{0k}_{bc} \) -\Lambda \sqrt{h} \chi_i,
\label{Geq2}\\
e \bG^a_0 &=& \frac14\,\eps_{ijk} \eps^{abc} \( N^d E^i_d F^{jk}_{bc} - 2 E^i_b F^{jk}_{0c} \) -\Lambda \sqrt{h} N^a,
\label{Geq3}\\
e \bG^a_i &=&
-\frac14\,\eps_{ijk} \eps^{abc} \( \(\CN+N^d E_d^l\chi_l\) F^{jk}_{bc} - 2 E^j_d N^d F^{0k}_{bc}
- 2 E_b^l \chi_l F^{jk}_{0c} + 4 E^j_b F^{0k}_{0c}\)
\nonumber\\
&& +\Lambda \sqrt{h}\(\CN E^a_i +N^a\chi_i\).
\label{Geq4}
\eeq
\end{subequations}
It is easy to see that equations \eqref{Geq1} and \eqref{Geq2} are independent of time derivatives and the variables
playing the role of Lagrange multipliers. Thus, in the canonical formulation they will correspond to the four constraints
responsible for the diffeomorphism symmetry.

Furthermore, let us assume that the torsion-less condition has been solved,
so that the Einstein tensor $\eG^{\m\n} = \eG^\m_I e^{I\n}$ is symmetric. Its ten components
can be then conveniently projected along the time-like and space-like sheets using the $2+2$ formalism.
As is well-known \cite{d'Inverno:1980zz,Torre:1985rw,Goldberg:1992st,d'Inverno:2006dy},
among the projected equations
one can identify a trivial equation, immediately satisfied as a consequence of the gauge fixing \Ref{lf-gauge},
three subsidiary equations holding everywhere provided they hold on a given hypersurface,
and two dynamical equations. The latter provide the dynamics for the conformal metric of
the two-dimensional surface $\Ss$, which carries the physical degrees
of freedom of gravity in this formalism. These two equations are denoted $\perp {\tilde \bG}^{ab}$
in the literature, meaning the traceless part of the projection on $\Ss$.
Although in the following we will not use all this machinery, we do need
the two dynamical equations which, in terms of our tetrad \Ref{docomptet},
can be shown to be given by
\be\label{dynEqs}
 \perp \tilde{\bG}^{ab} = \Pi^{ab}_{cd} \Bigl[\bG^{cd}+N^d \bG^{c0}+N^{c} \bG^{0d}+N^c N^d \bG^{00}\Bigr],
\ee
where
\be
\Pi^{ab}_{cd} := \gind^{a}_{(c}\gind^{b}_{d)}-\hf\, \gind^{ab} \gind_{cd}
\label{projPi}
\ee
is the traceless part of the projector on $\Ss$ defined on symmetric tensors.
This projector will play an important role also in our story distinguishing the sector where
the light front condition affects the canonical structure.

\section{Canonical analysis}
\label{sec-canonical}

In this section we present the canonical analysis of the first order formulation of general relativity on a null foliation.
Our starting point is the Hilbert-Palatini action \eqref{HPaction} where the tetrad is taken to be in the form \eqref{docomptet}.
Thus, as in the usual ADM analysis \cite{Arnowitt:1960es}, instead of $e_\mu^I$, our dynamical variables will be $E_a^i$, $\CN$, $N^a$, $\chi^i$
and the components of the spin connection. However, one can take few shortcuts which streamline the analysis. We will now describe these
shortcuts and simultaneously outline the resulting canonical structure without going into calculational details presented
in the following subsections.
\begin{itemize}
\item
Since the lapse $\CN$, shift $N^a$ and time components of the spin connection $\omega_0^{IJ}$ appear in the action only linearly and
without time derivatives, one can exclude them from the phase space and consider as Lagrange multipliers from the very beginning.
The corresponding primary constraints, which we denote by $\CD_a$, $\CH$ and $\CG_{IJ}$, respectively, generate gauge symmetries of the theory
consisting of spacetime diffeomorphisms and local Lorentz rotations in the tangent space.

\item
Other components of the tetrad also enter the action without time derivatives. But instead of imposing the constraints that their momenta
vanish, one can profit the fact that one works in the first order formalism and use them to construct the momenta for $\omega_a^{IJ}$.
However, since $\omega_a^{IJ}$ have 18 components, whereas $E_a^i$, $\chi^i$ provide only 12, the resulting momenta,
which we denote $\tP^a_{IJ}$, have to satisfy 6 constraints $\Phi^{ab}$.
These are primary constraints to be added to the Hamiltonian description of the action.
They are quadratic in $\tP^a_{IJ}$, and are referred to as {\it simplicity constraints},
as they imply that, as an internal 2-form, $\tP^a_{IJ}$ is simple.\footnote{They play a prominent role
in the construction of spin foam models of quantum gravity, where they coincide with
the spatial part of the discrete covariant simplicity constraints \cite{Barrett:1997gw,Freidel:1998pt,Alexandrov:2008da}.}

\item
It is often convenient to break a part of the internal gauge symmetry and treat from the start $\chi^i$
as a fixed vector.\footnote{Such a gauge fixing is fine also at quantum level
and it is needed anyway once one tries to quantize the theory via,
for instance, the path integral technique \cite{Alexandrov:2012pj}.}
In particular, this fixes uniquely the type of foliation, as seen above, and in such setup
the light front condition \Ref{lfcond}
is just a condition on the gauge fixing. If we do so, we lose 3 independent momenta, namely,
the gauge fixing of $\chi^i$ gives rise to 3 additional constraints on the momenta
conjugate to $\omega_a^{IJ}$. Combining them with $\Phi^{ab}$, one arrives at nine constraints,
which is nothing else but simplicity constraints in their linear form \cite{Engle:2007uq,Alexandrov:2007pq},
and can be conveniently written as $\Phi_I^a$ with $x^I\Phi^a_I=0$.
\end{itemize}

Thus one arrives at the following picture. Having fixed the variables $\chi^i$, first order gravity can be formulated
on the $2\times 18$ dimensional phase space with $3+1+6+9=19$ primary constraints.
Then, following Dirac's algorithm, one has to study the stability of the constraints under time evolution.
The analysis turns out to be significantly different in the null case than in the previously treated space-like and time-like cases,
which is controlled by the norm of $\chi^i$.

If $\chi^2\ne 1$, the stabilization of the primary constraints leads to 6 secondary constraints, denoted $\Psi^{ab}$,
forming second class pairs with 6 of the primary simplicity constraints \cite{Peldan:1993hi}.
Furthermore, 3 of the Gauss constraints $\CG_{IJ}$ do not commute with the remaining primary simplicity and become second class,
consistently with the fact that conditions on $\chi^i$ gauge fix 3 boosts in the tangent space.
The situation is summarized by the following scheme, where
the arrows indicate which and how many constraints are mutually non-commuting:
\begin{center}
\begin{tabular}{cccccc}
primary constraints & $\Phi^a_I$ & $\stackrel{3}{\leftrightarrow}$ & $\CG_{IJ}$ & $\CD_a$ & $\CH$
\\
& $\updownarrow\scriptstyle{6}$ & & & &
\\
secondary constraints & $\Psi^{ab}$ & & & &
\end{tabular}
\end{center}
As a result, one has 7 first class and 18 second class constraints leaving behind the 4-dimensional phase space, as it should be
for 2 physical degrees of freedom.

If $\chi^2=1$, the stability analysis is quite different.
As will be shown below, one again finds 6 secondary constraints $\Psi^{ab}$,
but their stabilisation now leads to two further, tertiary constraints, which we denote $\Upsilon^{ab}$.
(They satisfy certain projection condition which leaves only 2 independent components.)
The structure of non-vanishing commutators also changes and leads to the following diagram:
\vspace{0.4cm}
\begin{center}
\begin{tabular}{cccccc}
primary constraints & $\Phi^a_I$ & $\stackrel{2}{\leftrightarrow}$ & $\CG_{IJ}$ & $\CD_a$ & $\CH$
\\
& $\updownarrow\scriptstyle{4}$ & & & &
\\
secondary constraints & $\Psi^{ab}$ & & & &
\\
& $\displaystyle{\circlearrowright\scriptstyle{2}}\atop \vphantom{a}$ & & & &
\\
tertiary constraints & $\Upsilon^{ab}$& & & &
\end{tabular}
\end{center}

\vspace{-3.7cm}\hspace{5.75cm}
\unitlength 0.44mm 
\linethickness{0.4pt}
\ifx\plotpoint\undefined\newsavebox{\plotpoint}\fi 
\begin{picture}(110,74)(0,0)
\put(57,7){\vector(-3,-4){.09}}\put(57,49){\vector(-3,4){.09}}\qbezier(57,49)(73,28)(57,7)
\put(130,60){\vector(2,-1){.09}}\put(50,60){\vector(-2,-1){.09}}\qbezier(50,60)(90,74)(130,60)
\put(69,30){\makebox(0,0)[cc]{\scriptsize 2}}
\put(90,71){\makebox(0,0)[cc]{\scriptsize 1}}
\end{picture}\label{page-diag}
\vspace{0.3cm}
\\
Comparing to the space-like case, one can note the following differences:
\begin{itemize}
\item
Only 2 of the Gauss constraints $\CG_{IJ}$ do not commute with the primary simplicity.
Hence only 2 boosts are used to gauge fix $\chi^i$.
The third component, the norm of $\chi^i$, provides a gauge fixing condition for
the Hamiltonian constraint which now becomes second class. 4 Gauss constraints remain first class,
consistently with the 4 generators of the Lorentz group preserving the hypersurface geometry.
\item
Similarly, only 4 of the secondary constraints do not commute with the primary simplicity.
The remaining 2 constraints turn out to be mutually non-commuting.
\item
The new tertiary constraints do not commute with those primary simplicity which previously did not commute with $\Psi^{ab}$.
\end{itemize}
Altogether, this gives 7 first class and 20 second class constraints, and the symplectic reduction produces a 2-dimensional phase space.
As we explained in section \ref{sec-LFfeatures}, this is precisely what one needs to describe two degrees of freedom on the light front.
Concerning the geometric interpretation of the constraints, we notice that, as in the complex self-dual formulation of \cite{Goldberg:1992st}, 
the first class part of the algebra is a genuine Lie algebra, given by the semi-direct product of the spatial diffeomorphisms 
and the internal gauge group associated with the isometries of a null hyperplane. In particular, the hypersurface diffeomorphisms 
have a particularly simple form, unlike in the metric case (cf. \cite{Torre:1985rw}).

In the rest of this section we provide the details leading to the above picture.
We will use a Lorentz covariant notations despite a part of this symmetry
is explicitly broken by the gauge choice of $\chi^i$.
This allows to write down all equations in a concise form
and to keep them as close as possible to the space-like case.
A non-covariant description, where the direction identified by $\chi^i$
is explicitly singled out can also be useful, as it provides a more geometric
insight into the role played by different components of the tetrad and the connection.
We report the main features of such analysis in section \ref{subsec-noncov}.

\subsection{Hamiltonian}
\label{subsec-Ham}

As a first step, we put the action \Ref{HPaction} in Hamiltonian form.
Plugging the decomposition \eqref{docomptet}, \eqref{deflapse} into \eqref{HPaction} and integrating by parts, we get
\be
S=
\int_\mathcal{M}\de^4\xc\Biggl[\tP^a_{IJ}\p_t \omega_a^{IJ}+\omega_0^{IJ} D_a\tP^a_{IJ}
-N^a\tP^b_{IJ} F_{ab}^{IJ}+\tNn\(\hf\,\tp^a_I\tp^b_J F_{ab}^{IJ}-\Lambda h\)\Biggr],
\label{action}
\ee
where we defined
\beq
\tP^a_{IJ}&=&\frac{1}{4}\, \eps^{abc}\eps_{IJKL} e_b^K e_c^L=\tp^a_{[I}\xpd{J]},
\label{defP}
\\
\tp^a_I &=& \(0,\tE^a_i\).
\label{noncov-field}
\eeq
The field $\tp^a_I$ here introduced is not covariant, we will comment on this in a moment.
The fields $\omega_0^{IJ}$, $N^a$ and $\tNn$
appear linearly and without time derivatives and therefore play the role of the Lagrange multipliers.
The phase space is parametrized by $\omega^{IJ}_a$ and $\tP^a_{IJ}$ with the symplectic structure given by
\be
\left\{\omega^{IJ}_a(\xc),\tP^{b}_{KL}(\yc)\right\}=\delta^b_a\delta^{IJ}_{KL}\delta^3(\xc,\yc).
\ee
The momenta $\tP^a_{IJ}$ are constructed from the triad $E_a^i$ and the vector $\chi^i$ which, according to our strategy
explained above, is considered to be non-dynamical. Due to the mismatch between the number of components,
$\tP^a_{IJ}$ should satisfy 9 constraints. Indeed, it is easy to see that \eqref{defP} implies
\be
{\eps_{IJ}}^{KL}\tP^a_{KL}\xp^J=0,
\label{constr-P}
\ee
whereas contraction of this equation with $\xp^I$ vanishes identically for any $\tP^a_{IJ}$.

This analysis makes it clear that there are 4 sets of primary constraints imposed on the kinematical phase space
\be
\begin{split}
\CG_{IJ}:= &\, D_a\tP^a_{IJ}=0,
\\
\CC_a:=&\,-\tP^b_{IJ} F_{ab}^{IJ}=0,
\\
\CH:=&\, \hf\,\tp^a_I\tp^b_J \(F_{ab}^{IJ}-\frac{\Lambda}{3}\, \eps_{abc}{\eps^{IJKL}}\xp^K \tp^c_L\)=0,
\\
\Phi^a_I:=&\, {\eps_{IJ}}^{KL}\tP^a_{KL}\xp^J=0.
\end{split}
\label{primconstr}
\ee
As anticipated above, the Hamiltonian constraint is written in terms of the non-covariant field $\tp^a_I$.
In fact, it cannot be written using the covariant $\tP^a_{IJ}$ in a direct way.
In order to do that, we have to introduce the unit time-like vector $\tau^I=(1,0)=\hf(\xp^I-\xm^I)$, which allows us to write
\be
\tp^a_I=- 2\tP^a_{IJ}\tau^J.
\label{expr-tp}
\ee
In other words, to write the constraint, we have to project out the $0$ component of the canonical field,
which can only be achieved including both null vectors, $\xp^I$ and $\xm^I$.
This is the price one should pay for describing the null foliation in the covariant framework and
is related to the lack of a canonical choice for the shift vector.

As in the space-like case, it is convenient to redefine the constraint $\CC_a$ to
\be
\CD_a:=\CC_a+\omega_a^{IJ}\CG_{IJ}=\p_b\(\tP^b_{IJ}\omega_a^{IJ}\)-\tP^b_{IJ}\p_a\omega_b^{IJ},
\label{diffconstr}
\ee
which turns out to be the true generator of 3-dimensional diffeomorphisms.
This in turn can be achieved by redefining in the action the Lagrange multiplier as
\be
\omega_0^{IJ}=n^{IJ}+N^a\omega_a^{IJ}.
\ee
Thus, using the standard notation for smeared constraints,
the total Hamiltonian generating the time evolution reads as
\be
-H_{\rm tot}=\CG(n)+\CD(N)+\CH(\tNn)+\Phi(\LMsim).
\label{Hamilt}
\ee

\subsection{Primary constraints}
\label{subsec-primary}

The next step is the analysis of the stability conditions for all primary constraints.
Since the Hamiltonian is a linear combination of these constraints, their time evolution follows from
their algebra, which can be found in Appendix C. As reported there in \eqref{alg-prime}, it turns out that
the only weakly non-vanishing commutators are those with the primary simplicity constraints $\Phi^a_I$.
They lead to the following stability conditions.

First, for the Gauss law we have\footnote{To show this, one needs to use the property
$$
\xp^{[I}{\eps^{J]KL}}_{M}\, \tp^a_L\xp^M=-\hf\,{\eps^{IJL}}_{M}\xp^M \(\delta^K_L(\tp^a\xp)-\tp^a_L\xp^K\).
$$
}
\be
\dot\CG_{IJ}=\{\CG_{IJ},H_{\rm tot}\}\approx -\hf\,\eps_{IJKL}\xp^L\((\tp^a \xp)\LMsim_a^K-\tp^{a,K}(\xp\LMsim_a)\)=0.
\label{stabCG}
\ee
Taking into account that the Lagrange multiplier $\LMsim_a^I$ by definition does not have components along $\xp^I$
(i.e. it can be chosen to satisfy $(\LMsim_a\xm)=0$), it can be algebraically decomposed as follows,
\be
\LMsim_a^I=\(\eta^{IJ}-\hf\,\xp^I\xm^J\)\(\lambda_{ab}\tp^b_J+\eps_{abc}\tp^b_J \kappa^c\),
\label{reprlambda}
\ee
where 6 components of the symmetric matrix $\lambda_{ab}$ and 3 components of $\kappa^a$ encode 9 independent components of $\LMsim_a^I$.
Substituting \eqref{reprlambda} into \eqref{stabCG}, one finds a condition on $\kappa^a$
\be
\eps^{IJKL}\xp^K\tp^a_L\eps_{abc}(\tp^b\xp)\kappa^c=0,
\label{stabkappa}
\ee
which is solved by $\kappa^a=\ukap(\tp^a\xp)$, where $\kappa$ is an arbitrary function.
Thus, the stability condition fixes two components of $\LMsim_a^I$ which, as a result, takes the form
\be
\LMsim_a^I=\(\eta^{IJ}-\hf\,\xp^I\xm^J\)\(\lambda_{ab}\tp^b_J+\ukap\eps_{abc}\tp^b_J (\tp^c\xp)\).
\label{reprlambdafree}
\ee

The next constraint to consider is $\CD_a$.
Its stability condition reads
\be
\dot\CD_a=\{\CD_a,H_{\rm tot}\}\approx \eps_{IJKL}\LMsim_b^I\tp^{b,J}\xp^K\p_a\xp^L=0,
\label{stabCD}
\ee
which is again an equation on $\LMsim_a^I$. However, it is identically satisfied upon substitution of \eqref{reprlambdafree}.
Thus, the stability of the diffeomorphism constraint does not impose any new conditions.

Now we turn to the simplicity constraint. Its stability condition gets contributions from all commutators
and is given by
\be
\begin{split}
\dot\Phi^a_I=\{ \Phi^a_I,H_{\rm tot}\}=&\, \hf\,{\eps_{NJ}}^{KL}\xp^J n_{KL}\(\delta_I^N(\tp^a\xp)-\tp^{a,N}x_{+,I}\)
\\
&\, -N^b\eps_{IJKL}\tp^{a,J}\xp^K\p_b\xp^L
+{\eps_{IJ}}^{KL}\xp^J D_b\(\tNn \tp^a_K\tp^b_L\)=0.
\end{split}
\label{stabPhi}
\ee
To elucidate the content of this condition, let us
contract\footnote{Recall that $\xp^I \Phi^a_I \equiv 0$, thus this contraction does not lose any equation.}
it with $\tp^b_I$ and split the resulting tensor into symmetric and antisymmetric parts in the free indices $ab$.
The symmetric part is
\be
\tNn {\eps_{I}}^{JKL}\xp^I\tp^{(a}_J \tp^c_K D_c\tp^{b)}_L=0,
\label{seccon}
\ee
and since the lapse cannot be vanishing, this equation generates 6 secondary constraints.
The antisymmetric part reads
\be
\eps_{abc}{\eps_I}^{JKL}\xp^I\Bigl(\tp^b_J(\tp^c\xp)n_{KL}+N^d\tp^b_J\tp^{c}_K\p_d \xpd{L}
+\tp^b_J D_d\(\tNn \tp^c_K\tp^d_L\)\Bigr)=0,
\label{stabPhiantisym}
\ee
and can be further split. Contraction with $(\tp^a\xp)$ kills the first two terms,
and one remains with a scalar {\it differential} equation for the lapse function,
\be
\eps_{abc}(\tp^a\xp){\eps_{I}}^{JKL}\xp^I\tp^b_J D_d\(\tNn \tp^c_K\tp^d_L\)=0.
\label{stablpase}
\ee
The remaining 2 components of \eqref{stabPhiantisym} fix 2 components of the Lagrange multipliers
$n^{IJ}$ of the Gauss constraint. These equations can be easily identified with the decomposition
of Cartan's equations studied in section \ref{subsec-Cartan}, in particular \Ref{stablpase} coincides
on the constraint surface with the equation \eqref{constrN} for the lapse.

Finally, we have to analyze the stability of the Hamiltonian constraint $\CH$. It gives
\be
\dot \CH=\{\CH,H_{\rm tot}\}\approx {\eps_{IJ}}^{KL}\tp^a_K\tp^b_L D_b\(\xp^J\LMsim_a^I\)=0.
\ee
Substituting \eqref{reprlambdafree} for $\LMsim_a^I$, one obtains that the term with $\lambda_{ab}$ is proportional
to the secondary constraint $\Psi^{ab}$ so that the stability condition reduces to an equation fixing $\kappa$:
\be
\eps_{abc}{\eps_{I}}^{JKL}\tp^c_K\tp^d_L D_d\(\ukap\xp^I\tp^b_J(\tp^a\xp)\)
\approx 2\sqrt{h}\Bigl( \p_a\(\tE^a_i\chi^i\kappa\)+\kappa \tE^a_i\chi^i\omega_a^{0j}\chi_j\Bigr)=0.
\label{condalph}
\ee
Note that both \Ref{stablpase} and \Ref{condalph} are linear differential equations
with the differential operator given by $\p_-$. This is relevant for the analysis of zero modes in section \ref{subsec-diff}.

Thus, at this stage we fixed 6 Lagrange multipliers (the lapse $\tNn$, 2 components of $n^{IJ}$ and 3 components of $\LMsim_a^I$)
and generated 6 secondary constraints \eqref{seccon}. We now move to the next step and impose
 stability of the secondary constraints.

\subsection{Secondary constraints}
\label{subsec-secondary}

To stabilize the secondary constraints coming from \Ref{seccon}, let us first rewrite them
in terms of canonical variables. Using \eqref{defP} and \eqref{expr-tp}, this can be done as follows\footnote{Here we used
the symmetry properties of the indices to bring $x_+^I$ appearing in \Ref{seccon} inside the derivative,
and directly traded for $\tP^a_{IJ}$ using \eqref{defP}. Direct substitution of \eqref{expr-tp} leads to
a slightly different expression, and thus some different commutation relations.
However, the two constraints defined in this way differ by terms proportional to the simplicity constraints $\Phi^a_I$,
which have been already stabilized, and therefore lead to the same canonical structure.
\label{foot-ambig}}
\be
\Psi^{ab}:= 4\eps^{IJKL}\tP^{(a}_{IM}\tau^M\tP^c_{JN}\tau^N D_c\tP^{b)}_{KL}=0.
\label{psiinP}
\ee
The commutation relations of $\Psi^{ab}$ with other constraints can be found in \eqref{alg-secprim}.
As a result, the stability condition for the secondary constraints gets two non-vanishing contributions:
from the commutators with the primary simplicity and the Hamiltonian constraint.
Furthermore, upon substitution of \eqref{reprlambdafree}, the contribution proportional to $\kappa$ vanishes and
one remains with the following condition
\be
\CM^{ab,cd}\lambda_{cd}
-\frac{1}{2}\,\tNn \eps^{(acd}F_{cd}^{MN}\xmd{M}\tp^{b)}_N=0,
\label{stabpsi}
\ee
where we introduced the matrix
\be
\CM^{ab,cd}=\eps^{(acg}\eps^{b)df}\gind_{gf}
\label{matrM}
\ee
defined by the induced metric $\gind_{ab}$. The crucial feature of this matrix is that, being considered as an operator
on the space of symmetric tensors, it has a two-dimensional kernel.
Indeed, it satisfies the following property
\be
\CM^{ab,cd} \Pi_{cd}^{gf}=0,
\label{Mprop}
\ee
where $\Pi_{cd}^{ab}$ is the projector \Ref{projPi} on the two-dimensional subspace of symmetric tensors
which are traceless and orthogonal to $(\tp^a\xp)=(\tE^a\chi)$.
This property can be traced back to the degeneracy of the induced metric and is a direct consequence
of the light front condition.

Due to \eqref{Mprop}, the stability condition \eqref{stabpsi} can be split into two parts.
If one projects it using the projector orthogonal to $\Pi^{ab}_{cd}$, then one obtains an equation fixing $\lambda_{ab}$,
or more precisely its 4 components encoding the trace part and the part along
the null vector of the induced metric, that is $(\tE^{b}\chi)\lambda_{ab}$.
On the other hand, under the projection by $\Pi^{ab}_{cd}$ the first term vanishes and one finds that the stability condition
generates two tertiary constraints
\be
\begin{split}
\Upsilon^{ab}:=&\, \hf\, \Pi^{ab}_{cd}\,\eps^{(cgf}F_{gf}^{MN}\xmd{M}\tp^{d)}_N
\\
=&\,
\hf\,\Pi^{ab}_{cd} E^{(c}_i\eps^{d)gf}\( F^{0i}_{gf}- \chi_j F^{ij}_{gf}\)=0.
\end{split}
\label{tertcon}
\ee
As a consequence, the two components of the Lagrange multiplier $\lambda_{ab}$ singled out by the projector,  which we
denote $\hlam_{ab}=\Pi_{ab}^{cd}\hlam_{cd}$, remain free.

This is the main difference of the canonical analysis on the light front with the one done on a space-like foliation.
In that case the matrix $\CM^{ab,cd}$ is non-degenerate so that the stabilization of the secondary constraints fixes all Lagrange multipliers
of the primary simplicity and the analysis stops at this point. As we see, on the light front the situation is different
and we have to perform one step more by stabilising the new constraints \eqref{tertcon}.

\subsection{Tertiary constraints}
\label{subsec-tertiary}

To complete the analysis, we need to ensure the stability of the tertiary constraints.
The explicit form of the stability condition
is rather long due to the complicated form of the commutation relations between $\Upsilon^{ab}$
and the primary constraints, but it is not necessary for our purposes.
Indeed, if the stabilization of $\Upsilon^{ab}$ does not generate any further constraints, the stability condition must
fix the two components of the Lagrange multiplier $\hat\l_{ab}$ which have remained free up to now.
Thus, it is sufficient to prove that the equation of the form
\be
\left\{\Upsilon^{ab},\int \de^3 \xc\, \eta^{IJ}\hlam_{cd}\tp^c_I\Phi^d_J\right\}=\cdots
\ee
is solvable with respect to $\hlam_{ab}$. Evaluating the Poisson bracket, one finds
\be
\Pi^{ab}_{cd}\,\eps^{cgf}{\eps^{IJ}}_{KL}\tp^{d}_I\xm^L \CD_f\(\tp^r_J\xp^K\hlam_{gr}\).
\ee
Let us concentrate on the terms where the derivative hits $\hlam_{gr}$. Using the properties of the projector,
these terms can be simplified to
\be
2h\Pi^{ab}_{cd}\,\eps^{cgf}\eps^{drp}(E_p\chi)\Pi_{gr}^{st}\p_f\hlam_{st}=-2\Pi^{ab,cd} \p_- \hlam_{cd}.
\ee
Thus, the stability condition takes the following schematic form
\be
\p_- \hlam_{ab}+ \CO\(\hlam\CD(\cdots)\)=\cdots,
\label{stabUps}
\ee
and indeed can be solved with respect to $\hlam_{ab}$, up to possible zero modes of the operator $\p_-$,
as typical for light front field theories.

\bigskip

This result ends the stabilization procedure. The constraints can be now classified into
first and second class either using their Poisson bracket algebra reported in Appendix C,
or looking at which Lagrange multipliers have been fixed and which have remained free.
The only non-trivial part of this classification concerns the secondary constraints $\Psi^{ab}$.
Since the matrix $\CM^{ab,cd}$ \eqref{matrM} has rank 4, only 4 of them do not commute with
the primary simplicity constraints, and are thus immediately second class.
The remaining 2 components commute with all primary constraints.
They may not commute with the tertiary constraints, but commutation can be achieved adding
an appropriate combination of the primary simplicity. However, it turns out that they are non-commuting themselves.
Indeed, let us extract from the commutator \eqref{alg-psi}
the part corresponding to these two components. This can be done by substitution of the smearing functions of the form
$\hmu_{ab}=\Pi_{ab}^{cd}\hmu_{cd}$.
Then the commutator becomes
\beq
\left\{\Psi(\hmu),\Psi(\hnu)\right\}&=&
-\int \de^3 x\Bigl[
h^{3/2} \Pi^{ab}_{cd}\,\eps^{cgf}\eps^{drp}(E_p\chi)\Pi_{gr}^{st}\(\hnu_{st}\p_f\hmu_{ab}-\hmu_{ab}\p_f\hnu_{st}\)+\CO\(\hmu\hnu\CD(\cdots)\)\Bigr]
\nonumber\\
&=&\int \de^3 x\Bigl[
\sqrt{h} \Pi^{ab,cd}\(\hnu_{cd}\p_-\hmu_{ab}-\hmu_{ab}\p_-\hnu_{cd}\)+\CO\(\hmu\hnu\CD(\cdots)\)\Bigr].
\label{alg-psiPi}
\eeq
As in \eqref{stabUps}, the Poisson bracket is given by a linear differential operator with the principal part given by $\p_-$.
Thus, up to zero modes, this operator is invertible.
As a result, all secondary constraints are second class and one arrives
at the diagram and the counting of the constraints presented in the beginning of this section on page \pageref{page-diag}.
It is also easy to verify from \eqref{alg-prime} that, as in \cite{Goldberg:1992st,d'Inverno:2006hr},
the first class part of the constraint algebra, represented by
the spatial diffeomorphisms and the four Lorentz transformations generating isometries of the null hypersurface,
form a Lie algebra with true structure constants.

Before we finish this section, note that the commutator \eqref{alg-psiPi} is analogous to the commutator
\eqref{lfcom} in scalar field theory on the light front. This shows that the two secondary constraints singled out by the projector $\Pi^{ab}_{cd}$
are the standard light front second class constraints appearing for the physical degrees of freedom, in perfect agreement
with the fact that graviton has two propagating modes.
This identification will be even more apparent in the non-covariant formulation discussed in the next subsection.

\subsection{Non-covariant analysis}
\label{subsec-noncov}

The canonical structure presented above and the role of different constraints, in particular,
become clearer if we give up the covariant formulation used so far, and introduce variables adapted
to the existence of a fixed direction $\chi^i$ in the tangent space. Then, instead of
parametrizing the phase space by the spatial components of the spin connection
and their conjugate momenta satisfying the simplicity constraints, we can solve these constraints explicitly and
diagonalize the resulting kinetic term. This gives a direct access to the interpretation of various components of the physical fields.

Our starting point is the same 3+1 decomposed action \eqref{action} where now we substitute
the explicit expression for the momenta $\tP^a_{IJ}$ given from \eqref{defP} by
\be
\tP^a_{IJ} =\left\{
\begin{array}{ll}
\scriptstyle{(IJ)=(0i)}\ :\ &\hf\,\tE^a_i ,
\\
\scriptstyle{(IJ)=(ij)}\ :\ &\tE^a_{[i}\chi_{j]} .
\end{array}
\right.
\label{exprP}
\ee
The kinetic term is then diagonalized by the same change of connection variables as in the space-like case \cite{Alexandrov:1998cu}
\be
\begin{split}
\omega_a^{0i}=&\,\eta_a^i-\omega_a^{ij}\chi_j,
\\
\omega_a^{ij}=&\,\eps^{ijk} \(r_{kl}+\hf\, \eps_{klm}\omega^m\)\Et_a^l.
\end{split}
\label{decomp-con}
\ee
Thus, we traded $\omega_a^{IJ}$ for $\eta_a^i$, $\omega^i$ and symmetric $r_{ij}$.
In terms of the new variables the kinetic term takes the canonical form
\be
\int\de^4 \xc\Bigl[\tE^a_i\p_t \eta_a^{i}+\chi_i\p_t\omega^i\Bigr],
\label{kinterm}
\ee
whereas the primary constraints \eqref{primconstr} and \eqref{diffconstr}
(except the simplicity which has been explicitly solved by \Ref{exprP})
are given by
\beq
\Gn_i&:=& \eps_{ijk} \CG^{jk}
\,=\p_a\(\eps_{ijk}\tE^a_j\chi^k\)-{\eps_{ij}}^k\eta_a^{j}\tE^a_k-{\eps_{ij}}^k \omega^j\chi_k,
\nonumber\\
\Ln_i&:=&\ 2\CG_{0i}\quad
=\p_a\tE^a_i+\(\tE^a_i\chi_j-\tE^a_j\chi_i\)\eta_a^{j}-\CX_{ij}\omega^j,
\nonumber\\
\CD_a &=& \p_b\(\eta_a^i\tE^b_i\)-\tE^b_i\p_a\eta_b^i+\omega^i\p_a\chi_i,
\nonumber\\
\CH &=&  \hf\,\tE^a_i\tE^b_j F^{ij}_{ab}-\Lambda h
\label{noncov-constr}\\
&=&
-\tE^a_i\p_a\omega^i-\hf\,h \omega^i\p_a(h^{-1}\tE^a_i)-\eps^{ijk}\tE^a_i\Et_b^l r_{kl}\p_a\tE^b_j
+\tE^a_i\tE^b_j\eta_{[a}^i\eta_{b]}^j
\nonumber\\
&&
-\hf\, \tE^a_i\omega^i\eta_a^j\chi_j-\hf\, \tE^a_i\eta_a^i \omega^j\chi_j
-\frac14\,\CX_{ij}\omega^i\omega^j-\eps^{ijk}\tE^a_i\chi_k r_{jl}\eta_a^l
\nonumber\\
&&
-\hf\, \CM^{ij,kl} r_{ij} r_{kl}
-\Lambda h.
\nonumber
\eeq
Writing down the Hamiltonian constraint, we used the matrix
\be
\CM^{ij,kl}=\eps^{(ikm}\eps^{j)ln}\CX_{mn},
\label{matrMij}
\ee
which is nothing else but the lift of the matrix \eqref{matrM} to the tangent space,
namely, $\CM^{ij,kl}=h^{-1} E_a^i E_b^j\CM^{ab,cd}E_c^k E_d^l$. Hence, for $\chi^2=1$ it also has a two-dimensional kernel
which will play a crucial role in the following analysis.

From the kinetic term \eqref{kinterm}, we see that $\tE^a_i$ is the momentum conjugate to $\eta_a^i$,
$\chi^i$ is the momentum for $\omega_i$, and $r_{ij} $ has vanishing momentum, which we denote $\pi^{ij}$.
Furthermore, we wish at this point to gauge-fix $\chi^i$. Since $\omega_i$ is a dynamical variable,
we keep its conjugate momentum as $\chi^i$, and instead introduce a gauge fixing function for this momentum which will be called $\chif^i$.
Thus, the list of primary constraints \eqref{noncov-constr} must be completed by
\be
\begin{split}
\Phi^{ij}:=&\, \pi^{ij}=0,
\\
\varphi^i:=&\,\chi^i-\chif^i=0,
\end{split}
\label{noncov-primconstr-ad}
\ee
where the gauge fixing function satisfies the condition $\chif^2=1$ so as to put the theory on the light front.

The canonical analysis goes precisely along the same lines as the covariant one, and we do not report the details here.
However, this non-covariant analysis shows the detailed mechanism of what changes on the light front,
thanks to the explicit appearance of $\CM^{ij,kl}$ in the Hamiltonian constraint.
For generic $\chi^i$, $\CH$ is quadratic in all $r_{ij}$, the components of the connection
having vanishing momenta. On the light front this is not true anymore due to the degeneracy of $\CM^{ij,kl}$,
and two components of $r_{ij}$ enter only linearly.
As a result, the secondary constraints, obtained as
\be
\Psi^{ij}=\frac{\p\CH}{\p r_{ij}}=-\eps^{(ikl}\tE^a_k\Et_b^{j)} \p_a\tE^b_l+\eps^{(ikl}\tE^a_k\chi_l\eta_a^{j)}
-\CM^{ij,kl} r_{kl}
\label{noncovPsi}
\ee
and related to \eqref{psiinP} by contraction with the triad $\Psi^{ij}=h^{-1} E_a^i E_b^j\Psi^{ab}$,
do not depend on these two components. Thus, whereas for generic $\chi^i$ all of the $\Psi^{ij}$'s
can be solved with respect to $r_{ij}$, now the two constraints obtained by applying the projector on $\Ss$, that is
\be
\hat \Psi^{ij} = \Pi^{ij}_{kl} \Psi^{kl},
\label{projPsi}
\ee
should rather be considered as equations on $\eta_a^i$, which are the momenta for the physical
degrees of freedom of the metric. The two missing components of $r_{ij}$ are instead fixed using the tertiary constraints.
The two constraints $\hat \Psi^{ij}$ are in fact the gravity version
of the light front constraint \Ref{lfcon}. We see that they appear here as secondary constraints,
and not primary ones, as was in the example of the scalar field theory.
This is a direct consequence of having used a first order action.

Another important difference with respect to the space-like canonical analysis concerns the field conjugate to $\chi^2$.
From \eqref{kinterm}, it is clear that this is the component of the spin connection given by $\chi^i\omega_i$.
One can easily verify (see \eqref{noncov-constr}) that, precisely at $\chi^2=1$,
the only place where it appears is the Hamiltonian constraint.
This explains why this gauge corresponds to the gauge fixing of
the symmetry generated by $\CH$ and not the boosts,
as was the case for generic values of $\chi^i$.

\section{Peculiarities on the light front}
\label{sec-pecular}

In this section we collect and discuss various subtle issues arising in the light front formulation of the first order gravity,
which appear to be specific to the combination of the light front condition with the dynamical nature of spacetime.

\subsection{Origin of the tertiary constraints}
\label{subsec-origin-ter}

The most striking feature of the canonical analysis presented in the previous section is the presence
of the tertiary constraints $\Upsilon^{ab}$ \eqref{tertcon}.
It is natural to ask what Lagrangian equations of motion are described by these constraints.
Since they are expressed in terms of the curvature tensor, it is natural to expect that the constraints arise from Einstein's equations.
In fact, as we demonstrate in this section, they appear from a combination of Einstein's equations with Bianchi identities
\be
\bB^{\mu,I}:= \eps^{\mu\nu\rho\sigma}e_{\nu,J} F_{\rho\sigma}^{IJ}=0.
\label{Bianchi}
\ee

First, let us perform the 3+1 decomposition of the Bianchi identities
\begin{subequations}
\beq
\bB^{0,0}&=& \eps^{abc}E_a^i F_{bc}^{0i},
\\
\bB^{0,i}&=&\eps^{abc}E_a^j \(F_{bc}^{ij}+\chi_j F_{bc}^{0i}\),
\\
\bB^{a,0}&=&\eps^{abc}\(2E_b^i F_{0c}^{0i}-N^d E_d^i F_{bc}^{0i}\),
\\
\bB^{a,i}&=&\eps^{abc}\(2E_b^j \(F_{0c}^{ij}+\chi_j F_{0c}^{0i}\)-N^d E_d^j\( F_{bc}^{ij}+\chi_j F_{bc}^{0i}\)-\CN F_{bc}^{0i}\).
\eeq
\end{subequations}
Then it is straightforward to check that
\be
\begin{split}
&
E^{(a}_i\(\bB^{b),i}+N^{b)}\bB^{0,i}\) -2e\,\eps^{ijk}E^{(a}_j\chi_k\( \bG^{b)}_i+N^{b)}\bG^0_i\)
\\
&\qquad
=-\CN E^{(a}_i\eps^{b)cd}\( F^{0i}_{cd}- \chi_j F^{ij}_{cd}\)
+2E^{(a}_i\eps^{b)cd}E_c^j \CX_{jk} \(F_{0d}^{ik}-N^g F_{gd}^{ik}\).
\end{split}
\label{combeq}
\ee
Furthermore, a simple manipulation shows that
\be
E^{(a}_i\eps^{b)cd}E_c^j \CX_{jk} \eps^{ikm}=\CM^{ab,cd}\Et_c^m.
\label{propEE}
\ee
Therefore, upon applying the projector $\Pi^{ab}_{cd}$ on the identity \eqref{combeq}, the last term vanishes,
whereas the first term on the r.h.s. gives precisely the tertiary constraints.
Thus, we conclude that
\be
\Upsilon^{ab}=\frac{1}{2\CN}\, \Pi^{ab}_{cd}\,E^{c}_i\Bigl[2e\,\eps^{ijk}\chi_j\( \bG^{d}_k+N^{d}\bG^0_k\)-\bB^{d,i}-N^{d}\bB^{0,i}\Bigr].
\ee

Furthermore, it turns out that the two Einstein equations described by the tertiary constraints are precisely the dynamical equations
\Ref{dynEqs}. Indeed, it is straightforward to show that for vanishing torsion
\be
\begin{split}
\Upsilon^{ab}
=&\, -2h\CN\Pi^{ab}_{cd}\eps_{gfr}g^{cf}g^{0r}\(\perp \tilde{\bG}^{dg}\).
\end{split}
\ee
Thus, the tertiary constraints of the first order formalism coincide with the propagating equations of the metric formalism.
The fact that dynamical equations become constraints is a feature of combining
the use of connection variables with a null foliation.
Heuristically, this happens because both the first order formalism and the choice of a null coordinate as time reduce
by one the degree of time derivatives in the field equations.
Technically, the crucial role of the light front condition manifests in the fact that one needs
to use the degeneracy of the matrix $\CM^{ab,cd}$ in order to cancel the last term in \eqref{combeq}.

\subsection{`Lost' equation}
\label{subsec-lost}

The distinguishing feature of gravity is the dynamical nature of spacetime. Therefore, as was
explained in section \ref{subsec-dectetrad}, studying a null foliation in general relativity
requires imposing the gauge condition $g^{00}=0$. Plugging this condition into
the action, one of Einstein's equations is apparently lost: the new action depends only on 9 variables and
the equation obtained by variation with respect to $g^{00}$ is clearly missing.
This issue was studied in \cite{Torre:1985rw} using the metric formalism, and in  \cite{Goldberg:1992st} using Ashtekar variables.
As a remedy, it was suggested to extend the phase space and simultaneously add a set of constraints which would
reintroduce by hand the missing equation. However, we did not consider such ad hoc modifications in our paper,
and yet our canonical analysis reproduces all field equations. The reason for this automatic consistency
lies in the use of a first order action, as we now discuss.

Let us consider first the simpler case of a finite dimensional system. We assume that it has some gauge symmetry,
and an action that can be put in a first order form. Then, the crucial observation is that even if we eliminate
a variable through a certain gauge fixing, the action still depends on its conjugated variable.
As a result, the Hamiltonian formulation of the gauge fixed action is still based on the same phase
space as the original one. The only difference is that the gauge fixing condition converts
one of the original first class constraints into second class. Therefore, its Lagrange multiplier
is fixed by the stability procedure, and it is this key step that allows us to recover
the Lagrangian equation associated with the gauge-fixed variable. This mechanism is illustrated in details in Appendix \ref{ap-lost}.

For field theories, however, there is an additional complication that may arise.
Suppose that the Poisson bracket of the gauge fixing condition $\varphi$ with the gauge
fixed constraint $\Ggf$ produces a differential operator $\nabla$ with a non-trivial kernel,
\be
\{ \Ggf,\varphi(\mu)\}=\nabla \mu,
\qquad \exists\mu_0\ne 0:\ \nabla\mu_0=0.
\label{gfcomm}
\ee
Of course, this means that the gauge freedom generated by $\Ggf$ has not been completely fixed by the gauge condition.
In this situation, the missing equation can be recovered only up to the zero mode $\mu_0$.

This is precisely what happens in first order gravity on the light front.
In this case, the pair constraint/gauge-fixing is given by  the Hamiltonian constraint
and light front condition, so in the above notations, $\Ggf=\CH$ and $\varphi=\chi^2-1$.
To evaluate \Ref{gfcomm}, observe that the variable canonically conjugate
to $\chi^2$ is $\chi^i\omega_i$, see section \ref{subsec-noncov}.
Using the expression for the Hamiltonian constraint from \eqref{noncov-constr}, one finds
\be
\{\CH,\chi^2\}=
\frac{\d\CH}{\d(\chi\omega)}=-\sqrt{h}\[\p_- +\Bigl(\p_a(E^a_i\chi^i)+E^a_i\chi^i\eta_a^j\chi_j \Bigr)\]+\chi^i\CG_{0i}.
\label{oper}
\ee
Thus, one indeed obtains a linear differential operator which does have a non-trivial kernel.

Notice, however, that up to the last term, which vanishes on the constraint surface, one gets the \emph{same} differential operator which appears in equation \eqref{condalph} fixing the Lagrange multiplier $\tilde\kappa$. This should not be a surprise since $\tilde\kappa$ plays the role of $\mu$ in \eqref{gfcomm}.
Thus, the zero mode of the operator \eqref{oper} and the potentially missing part of Einstein's equations coincide with the zero mode
of this Lagrange multiplier. In the next subsection we argue that this zero mode should actually be forbidden by boundary conditions.
This means that $\kappa$ must be set to zero and all Einstein's equations follow from the canonical analysis.

\subsection{Residual diffeomorphisms and zero modes}
\label{subsec-diff}

So far our analysis was purely local. However, as discussed in section \ref{sec-LFfeatures},
an important part of the dynamics on the light front can be hidden in the sector of zero modes.
It is therefore relevant to ask whether this sector exists, and what role it plays if it does, in the case of general relativity.
However, in gravity the analysis of zero modes is made complicated by the highly non-linear dynamics, the generic appearance
of caustics and of spacetime singularities limiting the extent of the null sheet, and other phenomena
which manifest the geometric origin of the gravitational interaction.
Furthermore, the experience with field theories shows that zero modes are strongly affected by the choice
of boundary conditions, and in this paper we do not discuss this issue in detail. Nonetheless,
we would like to make a few general comments on the existence of zero modes in the first order
formalism presented here
and propose a preliminary analysis, mostly ignoring all these troublesome issues.

Our prime interest is to understand whether the infinite dimensional phase space derived so far
should be supplemented with a (measure zero) sector of zero modes undetermined by the initial conditions,
as it is the case for massless field theories on Minkowski spacetime, including linearized gravity.
To this end, we need to understand the constraint structure of this sector.
In particular, if the zero mode of some second class constraint turns out to be first class, this can signify that
the initial conditions may not fix the solution uniquely and additional data, typically encoded in the zero mode
of the corresponding Lagrange multiplier, should be taken into account.
Specifically, it may potentially happen for those constraints whose Lagrange multipliers
are determined by {\it differential} equations with the principal part given by $\p_-$, or, more generally,
whose second class nature follows from commutation relations involving this operator.
In our case we have 8 candidates which satisfy this criterium. Using the non-covariant notations of
section \ref{subsec-noncov} and lifting all objects to the tangent space using the triad, these 8 candidates are
\begin{itemize}
\item
the Hamiltonian constraint $\CH$;
\item
the primary constraint $\chi^i\varphi_i\equiv \varphi$ defined in \eqref{noncov-primconstr-ad}, which plays
the role of the light front condition;
\item
two primary constraints $\hat\Phi^{ij}=\Pi^{ij}_{kl}\Phi^{kl}$, two secondary constraints $\hat\Psi^{ij}$
and two tertiary constraints $\Upsilon^{ij}$.
\end{itemize}
Their zero modes require special attention because
the first two constraints have the commutation relation given by the differential operator \eqref{oper}
and the commutators of the other constraints are encoded in \eqref{stabUps} and \eqref{alg-psiPi} which have a similar form as well.

First, the appearance of the Hamiltonian constraint in this list has a simple interpretation:
the light front condition \eqref{lfcond}
or \eqref{lf-gauge}, realized canonically by the constraint $\varphi$,
does not completely fix time diffeomorphisms and there exists a residual gauge symmetry.
Indeed, an infinitesimal diffeomorphism transformation of $g^{00}$ is found to be
\be
\delta_\xi g^{00}=2g^{0\mu}\nabla_\mu \xi^0=-2\CN^{-1} \p_-\xi^0,
\label{difftr-g}
\ee
where we used \eqref{invmetric}. This result shows that the diffeomorphisms with the transformation parameter
satisfying $\p_-\xi^0=0$ leave the gauge \eqref{lf-gauge} invariant and appear as residual
gauge transformations.\footnote{Notice the apparent mismatch between this condition on $\xi^0$ and
the equation on the lapse \eqref{constrN}
due to the presence of a connection-dependent term in the latter.
The lapse is the Lagrange multiplier for the Hamiltonian constraint, which is usually associated with
the generator of time diffeomorphisms, and it might be tempting to identify it with $\xi^0$.
However, the correct generator of the Lagrangian symmetry \eqref{difftr-g} in the canonical formulation is given by
the total Hamiltonian \cite{Reisenberger:1995xh,Alexandrov:2001wt}
$$
\CD_0(\xi^0)=\int \de^3 \xc\, \xi^0 H_{\rm tot}.
$$
This means that the smearing function appearing in the generator in front of the Hamiltonian constraint is the product $\xi^0\CN$.
It is this function that should satisfy \eqref{constrN} and it does provided $\p_-\xi^0=0$ and $\CN$ fulfils \eqref{constrN}.}
Whether the residual transformations are an actual symmetry of the theory and the zero mode of $\CH$ is first class
depends on the concrete form of boundary conditions.
We postpone to future work a more detailed analysis of this issue.
In any case, such a zero mode would be a usual gauge symmetry which does not require specification
of any additional information beyond initial conditions.

Next, it is easy to see that the constraint $\varphi$ cannot have a first class zero mode.
If it were the case, it would generate a symmetry transformation, which shifts
one of the components of the spin connection, namely $\chi^i\omega_i$, and leaves
the other variables intact. However, such a symmetry would be in contradiction with
the Cartan equations which uniquely determine the connection in terms of the tetrad.
This implies that a non-vanishing solution of \eqref{condalph}
is inconsistent with any reasonable boundary conditions and the Lagrange multiplier $\kappa$
must vanish.

The most non-trivial is the set of constraints consisting of
$\hat\Phi^{ij}$, $\hat\Psi^{ij}$ and $\Upsilon^{ij}$.
It describes the dynamics in the sector corresponding to the physical gravitational modes
(transverse and traceless). In particular, as was noticed above, $\hat\Psi^{ij}$
are the standard light front constraints determining the momenta for the physical modes,
whereas $\Upsilon^{ij}$ in the metric formalism become the equations describing their propagation.
Thus, this is precisely the sector where zero modes are expected to appear.
Since these constraints form the chain ``primary$\to$secondary$\to$tertiary" in the stabilization
procedure, they must be simultaneously either first or second class. Furthermore, it is well known that
all constraints appearing as a result of the stabilization procedure of one first class constraint
realize the same Lagrangian gauge symmetry, which is generated at the canonical level by the sum of all these constraints
smeared with the same parameter, but differentiated by an increasing number of time derivatives.
Therefore, even if our zero modes turn out to be first class,
they realize not six, but only two Lagrangian gauge symmetries
corresponding to the following combined canonical generator\footnote{The appearance of the lapse in the arguments
of constraints is due to the way the secondary and tertiary constraints are defined.}
\be
\Upsilon(\epsilon_0)+\hat\Psi\bigl(\tNn^{-1}\p_+\epsilon_0\bigr)-\hat\Phi\Bigl(\tNn^{-1}\p_+\(\tNn^{-1}\p_+\epsilon_0\)\Bigr).
\ee
The existence of such symmetries depends on consistency of solutions of \eqref{stabUps} and \eqref{alg-psiPi}
with boundary conditions.
By analogy with the case of four-dimensional massless theories one may expect
that only global zero modes can arise here.
If this is the case, the parameter $\epsilon_0$ in the above generator can be a function of the light front time $\xcp$ only.

\section{Conclusions}
\label{sec-concl}

In this paper we extended the canonical analysis of general relativity on a null foliation to a
first order action in terms of real connection variables.
A characteristic feature of our analysis is the use a tetrad decomposition suitable for an arbitrary
foliation, whose nature (space-like, time-like or null) is encoded in the norm of an internal vector.
In particular, this allows to work with a non-degenerate triad
and be close as much as possible to the formalism underlying the loop approach to quantum gravity.
It is also possible to relate this formulation to the double-null or $2+2$ formalism of \cite{d'Inverno:1980zz},
which makes some geometric properties manifest, by using
the natural double null foliation carried by the local Minkowski metric of the fibre bundle.

The canonical structure of the theory is rather elaborated, with a stabilisation procedure that
stops only at the level of tertiary constraints, and a few novelties in the geometric meaning of
the constraints and their correspondence to Lagrangian equations.
In particular, the tertiary constraints turn out to originate from
the two Einstein's equations propagating the physical degrees of freedom.
This gives them the same status as the Hamiltonian constraint, to which they also have a resembling expression.

Finally, we provided a framework to discuss the issue of zero modes in gravity on the light front.
In particular, we showed that the existence of zero modes not captured by initial data on a null hypersurface
is related to the fact that some second class constraints have first class zero modes.
If this happens, the data to be added to initial conditions
are contained in the zero modes of the corresponding Lagrange multipliers.
In the case of gravity we identified the constraints where these effects can potentially appear.
Furthermore, we found that the standard light front conditions of the linearized theory
appear as a part of the secondary simplicity constraints, and discussed how boundary conditions affect
the existence of zero modes at non-perturbative level.

Given these results, there are many directions in which this work could be developed.
First, one can try to explicitly evaluate the Dirac brackets and formulate the dynamics on reduced phase space.
It would then be interesting to compare the resulting structure with the one
proposed in \cite{Reisenberger:2007pq,Reisenberger:2007ku,Reisenberger:2012zq}.
A related issue is to study the constraint-free data in our formulation.
Indeed, we have push forward in this paper the use of a single null hypersurface,
whereas the constraint-free data are typically described using two null hypersurfaces
and a space-like surface defined at their intersection. To that end it is also useful
to translate our results to the Newman-Penrose formalism. This can be easily done,
and for instance, Bondi's complex shear can be identified with a projection of $\eta^i_a$ on the space-like surface $\Ss$.

This relation may also be used to better understand the boundary conditions
to be imposed on our fields and their asymptotic properties.
In particular, our local analysis should be connected with the familiar notions of asymptotic flatness and data on future null infinity ${\cal I}^+$.
This will allow us to make contact with previous quantization attempts \cite{Ashtekar:1981sf,Frittelli:1996fu,Ashtekar:2014zsa},
but also with recent perturbative developments \cite{Strominger:2013jfa,Adamo:2014yya}. Furthermore,
the boundary conditions are crucial to determine the structure of the zero mode sector of the theory,
whose importance we have discussed at length in the main text.

Finally, our results can also be used to develop a dynamical treatment of null spin networks \cite{Speziale:2013ifa},
and as a starting point for spin foam models with null boundaries.

\section*{Acknowledgments}

We are grateful to C. Deffayet, T. Damour, P. Grang\'e, A. Neveu, S. Paston, A. Perez, R. Penrose, and E. Prokhvatilov
for valuable discussions. S.A. thanks CPT of Marseille for hospitality during several visits leading to this work.

\appendix

\section{Scalar field theory on the light front}
\label{ap-scalar}

Some features of the canonical structure of field theories on the light front are not easily found in the literature.
To illustrate some of the phenomena which played a role in the main text,
in particular the issue of zero modes and their treatment at the canonical level,
we take in this appendix the example of a scalar field theory. We split the discussion in two parts.
First, the two-dimensional massless case, which is special in many respects.
Then, the four-dimensional case.

\subsection{Free massless scalar in two dimensions}
\label{ap-free2d}

The two-dimensional massless scalar field represents one of the simplest field theories.
It is described by the wave equation
\be
\p_+\p_-\phi=0,
\ee
which is trivially solved in terms of two arbitrary functions of the light cone coordinates
\be
\phi=\phi_+(\xcp)+\phi_-(\xcm).
\label{solfreefield}
\ee
From the usual canonical point of view these two functions or, more precisely, their symmetric and antisymmetric combinations,
are related to the initial values of the field and its conjugate momentum, respectively. Our aim here is to understand how they
appear in the light front formulation of this trivial theory.

The starting point is the action in the coordinates \eqref{lfcoor}
\be
S=\int \de \xcp \de \xcm \, \p_+ \phi\p_- \phi .
\ee
Thinking about $\xc^+$ as a time coordinate, one arrives, as already mentioned in the main text, to the constraint
\be
\Psi:= \pi-\p_-\phi=0,
\label{lfcon-ap}
\ee
where $\pi$ is the momentum conjugate to $\phi$. As a result, the Hamiltonian in such formulation
is simply proportional to this constraint
\be
H=\int \de \xcm\, \lambda\Psi,
\label{scf-Ham}
\ee
where $\lambda$ is the corresponding Lagrange multiplier.
Using the canonical Poisson bracket
\be
\{\phi(\xcm),\pi(\yc^-)\}=\delta(\xcm,\yc^-),
\ee
the stability condition of $\Psi$ is found to be
\be
\p_+\Psi=\{\Psi,H\}=-2\p_-\lambda=0.
\ee
Thus, it partially fixes the Lagrange multiplier requiring that it is independent of the spatial coordinate $\xcm$
\be
\lambda=\lambda_0(\xcp).
\label{LMnotv}
\ee
This indicates that the zero mode of the constraint $\Psi$ coupled to the Lagrange multiplier $\lambda_0$
is first class, whereas the remaining part of the constraint is second class.
This is consistent with the Poisson bracket
\be
\left\{\Psi(\lambda),\Psi(\lambda')\right\}=\int \de \xcm \(\lambda' \p_-\lambda-\lambda\p_-\lambda'\)
\label{PhiPhi}
\ee
which is identical to the result \eqref{lfcom} presented in the main text.

From this analysis we conclude that the phase space of the light front theory is one-dimensional.
Thus, on the initial value surface we have to provide only the field itself, but not its conjugate momentum
which is fixed by the light front constraint $\Psi$.
These data can be clearly identified with the function $\phi_-(\xcm)$ in \eqref{solfreefield}.
But where is the second function $\phi_+$ hidden in this formalism?
As it turns out, it is encoded in the zero mode $\lambda_0$ of the Lagrange multiplier.

Indeed, the Hamiltonian equation of motion
\be
\p_+\phi=\{\phi,H\}=\lambda_0(\xcp)
\ee
identifies $\lambda_0$ with the derivative of $\phi_+$. Since $\lambda_0$ multiplies the first class part of the constraint,
which is given by
\be
\Psi_0:=\int \Psi\,\de\xcm=\int \pi\,\de\xcm,
\label{zeroPsi}
\ee
it is an arbitrary function of $\xcp$ which must supplement the initial conditions to fix a solution uniquely.
In this way the presence of the gauge symmetry realized by $\Psi_0$ allows to describe
the degrees of freedom not captured by the data on the null hypersurface.

One might wonder how this can be, given that the presence of a gauge symmetry usually implies a {\it reduction} of degrees of freedom.
In particular, specification of the corresponding Lagrange multiplier is usually interpreted just as a gauge fixing.
On the light front the situation is different due to a different physical interpretation of the quantities affected by the gauge symmetry.
Whereas for the standard gauge symmetry such quantities are considered as unobservable, in the case of the gauge symmetry generated by
the zero mode of the light front constraint this is not true. For instance, we know that the function $\phi_+$ in \eqref{solfreefield}
can be measured so that solutions differing by values of $\phi_+$ are physically inequivalent, despite it transforms under the action of
the first class constraint $\Psi_0$.
Thus, on the light front one should distinguish between the usual first class constraints and the ones describing
the physical zero modes.

\subsection{Scalar field in four dimensions}
\label{ap-scalar4d}

Let us now turn to the four-dimensional scalar theory allowing also for a non-vanishing mass and a non-trivial potential.
In the light cone coordinates, the action functional for such theory is given by
\be
S=\int \de \xcp \de \xcm \de^2 \xc^\bot\, \Bigl(\p_+ \phi\p_- \phi -\hf\,(\p_\bot\phi)^2 -\frac{m^2}{2}\, \phi^2-V(\phi)\Bigr).
\ee
It gives rise to the same light front constraint $\Psi$ \eqref{lfcon-ap} as in the 2d massless case
and with the same commutation relations \eqref{PhiPhi}.
The Hamiltonian however acquires now additional contributions due to the mass, the potential and the orthogonal dimensions
\be
H=\int \de \xcm\de^2 \xc^\bot\( \hf\,(\p_\bot\phi)^2+\frac{m^2}{2}\, \phi^2+V(\phi)+\lambda\Psi\).
\ee
This changes the stability condition of $\Psi$ which now becomes an {\it inhomogeneous} differential equation on the Lagrange multiplier
\be
\(\p_\bot^2-m^2\)\phi-V'(\phi)-2\p_-\lambda=0.
\label{eqlam-mass}
\ee
As in the previous subsection, it can be solved with respect to $\lambda$ up to its zero mode $\lambda_0(\xcp)$
which remains free. This indicates that the zero mode of the constraint \eqref{zeroPsi} again might be first class.
However, the inhomogeneity of the equation leads to new features. Integrating the stability condition \eqref{eqlam-mass}
over the whole line of $\xcm$, one kills the last term and remains with an equation which should be interpreted as
a new secondary constraint
\be
\zmcon:= \int \de \xcm\Bigl(\(\p_\bot^2-m^2\)\phi-V'(\phi)\Bigr)=0.
\label{zmconstr}
\ee
This is the so-called {\it zero mode constraint} \cite{Heinzl:1989nx,Yamawaki:1998cy} which determines the zero mode of
the scalar field in terms of the other modes. In the absence of the potential, it requires that this zero mode vanishes.
Otherwise, it can become non-trivial and give rise to various phenomena such as spontaneous symmetry breaking.

At the next step we compute
\be
\{ \zmcon,\Psi(\lambda)\}=\int \de \xcm\Bigl(\(\p_\bot^2-m^2\)-V''(\phi)\Bigr)\,\lambda.
\label{commPsiSigma}
\ee
In particular, one finds that $\zmcon$ does not commute with $\Psi_0$.\footnote{In fact, their commutator diverges. This divergence
arises because we compute a commutator between two conjugate modes of a continuum spectrum. A way to regularize it
is to put boundaries at finite $\xcm$, which eventually leads to the {\it discrete light cone quantization}
framework \cite{Maskawa:1975ky,Pauli:1985ps}.} Thus, they are both second class constraints and the stabilization of $\zmcon$,
which requires \eqref{commPsiSigma} to vanish, fixes the zero mode $\lambda_0$ of the Lagrange multiplier.
As a result, no first class constraints arise in this case and a solution of the theory is uniquely specified by the initial data
for the scalar field on the light front \cite{Heinzl:1993px}.
The only additional restriction is that these data should satisfy the zero mode constraint \eqref{zmconstr}.

Finally, we note that the case of a massless field in four dimensions represents a mixture of the structures
presented in this and the previous subsections. If we set $m^2=V=0$, one still gets the zero mode constraint \eqref{zmconstr}.
However, in contrast to the massive case, it satisfies
\be
\int \zmcon\,\de^2 \xc^\bot=0.
\ee
Thus, the constraint does not restrict the {\it global zero mode} which is independent of all spatial coordinates.
Similarly, from \eqref{commPsiSigma} one finds that $\zmcon$ commutes with
\be
\Psigl:=\int \Psi\,\de\xcm\de^2\xc^\bot,
\ee
which means that $\Psigl$ is first class and the global zero mode of $\lambda$ remains an undetermined function of $\xcp$,
which should be specified together with initial conditions.
This is, in fact, the expected result since in the massless case the global zero mode propagates parallel to the light front
as illustrated on Fig. \ref{fig-cone}.

\section{Decomposition of the metric}
\label{ap-metric}

In the light front gauge \eqref{lfcond}, the inverse tetrad can be found to be
\be
e^0_I=-e^{-1}\sqrt{h} \eta_{IJ} x^J,
\qquad
e^a_I=e^{-1}\sqrt{h}N^a \eta_{IJ} x^J+h^{-1/2}\tp^a_I,
\label{invtetrad}
\ee
where $\tp^a_I$ is defined in \eqref{noncov-field} and $e=\CN \sqrt{h}$ is the determinant of the tetrad.
The expressions for the metric and its inverse easily follow from \eqref{docomptet} and \eqref{invtetrad}, respectively,
\beq
g_{\mu\nu}&=&\(\begin{array}{cc}
-\CN^2 +N^a N^b \gind_{ab}-2\CN N^a E_a^i\chi_i & \gind_{bc}N^c -\CN E_b^i\chi_i
\\
\gind_{ac}N^c-\CN E_a^i\chi_i & \gind_{ab}
\end{array}\),
\label{metric}
\\
g^{\mu\nu}&=& \f1{\CN} \(\begin{array}{cc}
0 & - E^b_i\chi^i
\\
- E^a_i\chi^i & \CN E^a_i E^b_i+ (N^aE^b_i+N^bE^a_i)\chi_i
\end{array}\).
\label{invmetric}
\eeq

\section{Constraint algebra}
\label{ap-algebra}

The commutators of the primary constraints on the surface of the simplicity constraint $\Phi^a_I$ are given by
\beq
\{ \CG(n),\CG(m)\} &=& \CG(n\times m),
\nonumber\\
\{\CD(\vec N),\CG(n)\}&=&-\CG(N^a\p_a n),
\nonumber\\
\{\CD(\vec N),\CD(\vec M)\}&=&-\CD([\vec N,\vec M]),
\nonumber\\
\{\CG(n),\CH(\tN)\}&=& \CH\(\tN n_{IJ}\xp^I \xm^J\)+\CD(\vec U)+\CG(U^a\omega_a),
\nonumber\\
\{\CD(\vec N),\CH(\tN)\}&=&-\CH\(\CL_{\vec N}\tN\),
\label{alg-prime}\\
\{\CH(\tN),\CH(\tM)\}&=&0,
\nonumber\\
\{\CG(n),\Phi^a_I\}&=&{\eps_{IJ}}^{KL}\xp^J n_{KN}\xp^N \tp^a_L,
\nonumber\\
\{ \CD(\vec N),\Phi^a_I\} &=& -N^b\eps_{IJKL}\tp^{a,J}\xp^K\p_b \xp^L,
\nonumber\\
\{\CH(\tN),\Phi^a_I\}&=&{\eps_{IJ}}^{KL}\xp^J D_b\(\tN\tp^a_K\tp^b_L\),
\nonumber
\eeq
where
\be
\begin{split}
\(n\times m\)^{IJ}=&\, {n^I}_K m^{KJ}-{n^J}_K m^{KI},
\\
{[\vec N,\vec M]}^a=&\, N^b\p_b M^a-M^b\p_b N^a,
\\
\CL_{\vec N}\tN=&\,N^a\p_a \tN-\tN\p_a N^a,
\\
U^a =&\, -\tN n^{IJ}\tau_I\tp_J^a=\tN n^{0i}\tE^a_i.
\end{split}
\ee
The secondary constraints commute with the primary ones as follows
\be
\begin{split}
\{\Phi(\LMsim),\Psi^{ab}\}=&\,(\tp^{(a}\LMsim_c)(\tp^{b)}\xp)(\tp^c\xp)+(\tp^{(a}\xp)(\tp^{b)}\tp^c)(\LMsim_c \xp)
\\
&\,
-(\tp^{(a}\xp)(\tp^{b)}\xp)(\tp^c\LMsim_c )-(\tp^{a}\tp^{b})(\tp^c\xp)(\LMsim_c \xp),
\\
\{\CG(n),\Psi^{ab}\}=&\, -n_{IJ}\xp^I\xm^J\Psi^{ab}
- \(n^{IJ}\tau_I\tp^{(a}_J\){\eps_{K}}^{LMN}\xp^K\tp^{b)}_L \CG_{MN},
\\
\{\CD(\vec N),\Psi^{ab}\}=&\, N^d\p_d\Psi^{ab}+3\p_d N^d \Psi^{ab}-\p_d N^a\Psi^{db}-\p_d N^b \Psi^{ad},
\\
\{\CH(\tN),\Psi^{ab}\}=&\, 3\tN\(\tp^c_I D_c\tau^I\)\Psi^{ab}-2\tN\(\tp^{(a}_I D_c \tau^I\)\Psi^{b)c}
-2\tN \eps^{IJKL}\tp^{(a}_I\tp^c_J D_c\tp^{b)}_K \CG_{LM}\tau^M
\\
&\, +\frac{h}{2}\,\tN \eps^{(acd}F_{cd}^{MN}\xmd{M}\tp^{b)}_N,
\end{split}
\label{alg-secprim}
\ee
whereas their mutual commutator reads
\be
\{\Psi(\mu),\Psi(\nu)\}=\int \de^3 x\, {\eps^{IJ}}_{KL}\xp^K \eps^{I'J'K'L}\tp^g_{K'}
\Bigl[\(\nu_{cd}\tp^c_J\tp^d_{J'}\)D_g\(\mu_{ab}\tp^a_I\tp^b_{I'}\)-\(\mu_{ab}\tp^a_I\tp^b_{I'}\)D_g\(\nu_{cd}\tp^c_J\tp^d_{J'}\)\Bigr].
\label{alg-psi}
\ee

\section{Gauge fixing and missing equations}
\label{ap-lost}

In the main text we raised the issue that a gauge fixing leads to an apparent loss of one of equations of motion.
In this appendix we show that when one works with a first order action, the apparently missing equation is recovered from the stabilisation procedure.
Without loss of generality, we can assume that the system has the following structure
\be
S[q,p,\omega,\chi]=\int \de t\Bigl(p_i \p_t q^i +\chi\p_t\omega -H_T\Bigr),
\qquad
H_T=H(q,p,\omega,\chi)+\lambda^\alpha \Gg_\alpha(q,p,\omega,\chi),
\label{inact}
\ee
where $(q_i,p^i,\omega,\chi)$ span the phase space, $H$ is a Hamiltonian, and $\Gg_\alpha$ represent
a set of first class constraints.
We distinguished a pair of conjugate variables $(\omega,\chi)$ because our aim is
to investigate the difference between the system \eqref{inact} and the one obtained by a gauge fixing of the variable $\chi$.
Namely, let us assume that the condition $\chi=\chif(q,p,\omega)$ fixes the gauge symmetry generated
by one of the first class constraints, say $\Ggf$.
Then the gauge fixed action becomes
\be
S^{\rm g.f.}[q,p,\omega]=
\int \de t\Bigl(p_i \p_t q^i +\chif(q,p,\omega)\p_t\omega - H^{\rm g.f.}(q,p,\omega)-\lambda^\alpha \Gg^{\rm g.f.}_\alpha(q,p,\omega)\Bigr),
\label{Sgf}
\ee
where
\be
\begin{split}
H^{\rm g.f.}(q,p,\omega)=&\, H(q,p,\omega,\chif(q,p,\omega)),
\\
\Gg_\alpha^{\rm g.f.}(q,p,\omega)=&\, \Gg_\alpha(q,p,\omega,\chif(q,p,\omega)).
\end{split}
\ee
Since the gauge fixed action does not depend on $\chi$ anymore, we seem to lose one equation of motion
of the original system
\be
\frac{\delta S}{\delta\chi}=\p_t\omega-\p_\chi H_T=0.
\label{lost}
\ee
What is the fate of this equation in the gauge fixed theory?

To understand this issue, one should proceed with the Hamiltonian analysis of \eqref{Sgf}. Then, in addition
to the constraints $\Gg_\alpha^{\rm g.f.}$, one finds another primary constraint
\be
\varphi=\chi-\chif(q,p,\omega),
\ee
where $\chi$ is the momentum conjugate to $\omega$. This constraint should be added to the total Hamiltonian with a Lagrange multiplier $\mu$
\be
H_T^{\rm g.f.}=H^{\rm g.f.}+\lambda^\alpha \Gg^{\rm g.f.}_\alpha+\mu\varphi.
\ee
The next step is the stability analysis of all primary constraints. One finds
\beq
\p_t\varphi&=& \{\varphi, H_T^{\rm g.f.}\}
\nonumber\\
&=&\p_\omega H^{\rm g.f.}+\lambda^\alpha \p_\omega \Gg^{\rm g.f.}_\alpha
-\{\chif,H^{\rm g.f.}+\lambda^\alpha \Gg^{\rm g.f.}_\alpha\}
\nonumber\\
&=& \p_\omega H_T-\{\chif,H_T\}=0,
\label{stabphi}
\\
\p_t \Gg^{\rm g.f.}_\alpha&=& \{\Gg^{\rm g.f.}_\alpha, H_T^{\rm g.f.}\}
\nonumber\\
&=&
\{\Gg_\alpha,H_T\}^{\rm g.f.}-\p_\chi \Gg_\alpha\{\chif,H_T\}
+\{\Gg_\alpha,\chif\}\p_\chi H_T+\mu\(\p_\omega \Gg_\alpha-\{\Gg_\alpha,\chif\}\)
\nonumber\\
&\approx & \(\p_\omega \Gg_\alpha-\{\Gg_\alpha,\chif\}\)\(\mu-\p_\chi H_T\)=0,
\label{stabGalph}
\eeq
where we used the notation $\{f,g\}^{\rm g.f.}=\p_{q^i}f \p_{p_i}g-\p_{p_i}f \p_{q^i}g$, and to get the last line
we took into account \eqref{stabphi} and the stability of $\Gg_\alpha$ in the non-gauge-fixed theory,
which implies that $\{\Gg_\alpha,H_T\}\approx 0$.
Since the gauge condition is supposed to fix the symmetry generated by $\Ggf$, their commutator should be non-vanishing, i.e.
\be
\{\Ggf,\chi-\chif\}=\p_\omega \Ggf-\{\Ggf,\chif\}\ne 0.
\label{gfcond}
\ee
Then it is easy to see that the condition \eqref{stabphi} fixes the Lagrange multiplier $\lamgf$, whereas
the stability of all constraints $\Gg^{\rm g.f.}_\alpha$ is achieved by fixing the Lagrange multiplier $\mu$
\be
\mu=\p_\chi H_T.
\ee
It is this result that ensures the equivalence of the two systems because
the Hamiltonian equation of motion for $\omega$ in the gauge fixed case
\be
\p_t \omega=\{\omega, H_T^{\rm g.f.}\}=\mu=\p_\chi H_T
\ee
precisely coincides with the original equation \eqref{lost}. Thus, this equation is not lost, but it is still a part of the
partially gauge fixed canonical formulation.

\providecommand{\href}[2]{#2}\begingroup\raggedright\endgroup


\end{document}